\documentclass[oneside,a4paper,12pt,reqno]{amsart}
\usepackage{amsmath,amstext,amsbsy,amsopn,amsthm,amscd,amsfonts,amssymb}
\usepackage{graphicx}
\usepackage[round]{natbib}
\usepackage{times}
\usepackage{mathptmx}
\usepackage[colorlinks=true,citecolor=blue,linkcolor=blue]{hyperref}

\renewcommand{\baselinestretch}{1.45}

\newcommand{\begineq}[1]{\begin{equation}\label{#1}}
\newcommand{\refeq}[1]{Eq.\ (\ref{#1})}
\newcommand{\ie}{\emph{i.e.,}}
\newcommand{\eg}{\emph{e.g.,}}
\newcommand{\Eq}[1]{Eq.~(\ref{#1})}

\newcommand{\defgT}{\mathbf{F}}
\newcommand{\defg}{F}
\newcommand{\curposT}{\mathbf{x}}
\newcommand{\refposT}{\mathbf{X}}
\newcommand{\CGstrain}{\mathbf{C}}
\newcommand{\refnormalT}{\hat{\mathbf{N}}}
\newcommand{\cauchyT}{\boldsymbol{\sigma}}      
\newcommand{\cauchy}{\sigma}    
\newcommand {\nominalT}{\mathbf{P}}
\newcommand{\mapping}{\mathbf{\chi}}
\newcommand{\rotangle}{\beta}
\newcommand{\BI}[1]{\bar{I}_{#1}}
\newcommand {\volfrac}{c}
\newcommand{\Trace}{\text{Tr}}
\newcommand{\sedf}{\Psi}
\newcommand {\Rotation}{\mathbf{R}}

\newcommand {\mum}{\mu^{(m)}}
\newcommand {\muf}{\mu^{(f)}}

\begin{document}

\title[Analysis of the micromechanics of soft fibrous connective tissues]{Analytical and numerical analyses of the micromechanics of soft fibrous connective tissues}
\author{Gal deBotton$^{\dagger,\ddagger}$ and Tal Oren$^{\dagger}$}
\maketitle

\centerline{$^{\dagger}$Department of Biomedical Engineering, Ben-Gurion University}
\centerline{Beer-Sheva 84105, Israel}
\vspace{0.15in}
\centerline{$^{\ddagger}$The Pearlstone Center for Aeronautical Studies}
\centerline{Department of Mechanical Engineering, Ben-Gurion University}
\centerline{Beer-Sheva 84105, Israel}

\begin{abstract}
State of the art research and treatment of biological tissues require accurate and efficient methods for describing their mechanical properties.
Indeed, micromechanics motivated approaches provide a systematic method for elevating relevant data from the microscopic level to the macroscopic one.
In this work the mechanical responses of hyperelastic tissues with one and two families of collagen fibers 
are analyzed by application of a new variational estimate accounting for their histology and the behaviors of their constituents.
The resulting, close form expressions, are used to determine the overall response of the wall of a healthy human coronary artery. 
To demonstrate the accuracy of the proposed method these predictions are compared with corresponding 3-D finite element simulations of a periodic unit cell of the tissue with two families of fibers. 
Throughout, the analytical predictions for the highly nonlinear and anisotropic tissue are in agreement with the numerical simulations. 
\\
\textbf{keywords:} finite deformation; homogenization; variational estimate; soft tissue; arterial wall; collagen fiber; micromechanics; anisotropy
\\
\end{abstract}

\setcounter{footnote}{1}
\footnotetext{\lowercase{E-mail: debotton@bgumail.bgu.ac.il, Tel: int. (972) 8 - 647 7105,
Fax: int. (972) 8 - 647 7106}}

\section{Introduction}

Arteries and other blood vessels, tendons, ligaments, skin, cornea, cartilage, intervertebral discs and the gut, among others, are deformable collagenous tissues whose physiological functioning crucially depends on their mechanical properties.
Moreover, it is well known that the mechanical environment experienced by these tissues play a critical role in the origin and progression of diseases.
In particular, with respect to the cardiovascular system: stroke, aortic aneurisms and dissections, acute myocardial infraction (MI) and congestive heart failure (CHF) as well as restenosis following intravascular stenting and other cardiovascular diseases and injuries, result in significant morbidity and mortality worldwide. 
Accordingly, many methods of clinical treatment and surgical interventions largely rely on mechanical effects \citep{hump02book}. 
Over the last decade, the unique mechanical behavior of soft biological
tissues and particularly that of the arterial wall has been receiving considerable
attention in the literature. 
Despite the significant progress, there remains
a pressing need for a better understanding of the fundamental role
of mechanics in vascular physiology and pathophysiology, that will
eventually enable the prediction of the mechano-stimulated vascular
adaptation \citep{hump03prsla,holz&ogde10prsl}.

On top of the complications related to the biological aspects, even the characterization of the purely mechanical factors governing the response of soft connective tissues is not a straightforward procedure and requires a reliable constitutive model. 
The main constituents of soft tissues are: fibroblasts, the extra-fibrillar matrix (EFM) which includes the elastin, and the collagen fibers. 
From a mechanical viewpoint, the most important load-bearing component is the collagen fibers network.
The function, strength, integrity and stability of soft tissues are maintained by the structural arrangement of the collagen fibers.
Thus, changes in collagen content, orientation,
fiber type or fiber thickness, all have a major impact on the overall
responses of the tissues.
Many of the aspects related to the collagen fibers are discussed in \cite{fraz08book}.

The histology of collagenous tissues evolves in a way that optimizes their overall mechanical
functionality. Specifically, the mechanical behavior of the artery
depends on its anatomical site. For instance, \citet{tabe&hump01jbet} note
that the properties of the vascular tissues vary across the arterial
wall. 
Also, it has been demonstrated that the distribution of the angle of the collagen fibers network vary through the thickness of the artery, particularly at bifurcation sites \citep{holz&stad02annbioeng,har&etal07thebio}. 

Today it is widely accepted that changes in the ambient mechanical loads acting on the tissue are followed
by a remarkable continuous adaptation of the tissue.
The versatility of collagen as a building material is mainly thanks to
modification of its structure. Collagen adaptation to loading both
in healthy and diseased tissues involves two main processes: remodeling
and growth. Particularly, the fibers may grow and reorient in order
to meet new mechanical demands. Growth is defined as the change in
concentration due to mass transport. Remodeling is considered as reorganization
of existing fibers at constant mass \citep{ambr&etal11jmps}.

The list of examples of vascular growth and remodeling in normal development,
ageing, disease progression, functional adaptations to altered environments
and responses to injury or clinical treatment is endless \citep[\emph{e.g.},][]{ambr&etal11jmps}.
Hypertension is an example of a disease that causes
and is caused by alternations in the mechanical environment of the
arterial wall. In hypertension, the sustained increase in blood pressure
may result in a thickening of the wall as a consequence of collagen
deposition and associated restoration of the hoop stress
to its normal value. There are also changes in blood flow-induced
wall shear stress that effect the artery wall such as enlargement of lumen
diameter due to increased wall shear stress. Obviously, during the
development of hypertension, marked changes in the microstructure
and functionality of blood vessels take place. Growth and remodeling processes
of collagen fibers serve to restore the stresses to their homeostatic
levels.
These processes attempt to recover and maintain
the preferred form and function and are fundamental for adaptation to
altered blood flow \citep{hump08cellbiochembiophy,eber&etal10ajphcp}.
In the context of tissue engineering, \cite{nere&seli01arbe} emphasized the strong influence of mechanical loading on the remodeling of the collagen network. 

It is evident that in order to accurately capture the stress-strain response of the heterogeneous, anisotropic, and highly nonlinear tissue with its characteristic stiffening at large strains, it is necessary to develop a tool that enables to account for the specific contributions of the significant constituents to its overall mechanics. 
Moreover, to conduct remodeling analyses that provide further insight into the mechanical behavior of the phases composing the tissue, structurally-based constitutive models that incorporate collagen fiber alignments, are needed. 
The motivation for our research is best summarized in \cite{hump09jmmb} who states that one of the conspicuous shortcomings that have remained is 
``the need to account explicitly for separate contributions by the diverse structurally significant constituents within the arterial wall''.
Nonetheless, while phenomenological models that treat the arterial walls and other soft collagenous tissues as homogeneous materials were developed in recent years \citep[\emph{e.g.},][among others]{fung67ajph,lani83jbta,hump&yin87biophyj,holz&etal00jela}, to date only a few micromechanics-motivated studies exist in the literature \citep[\emph{e.g.},][]{gdb&shmu09jmps,chen&etal11jmps}.

The aim of this work is to introduce a novel comprehensive micromechanical-based
model for the determination of the macroscopic response of soft tissues
with two families of fibers in terms of the properties of the
constituting phases, their spatial distribution and the interactions
between them in a consistent way. 
At the present stage of the study we restrict our attention to passive response of tissues which are not pre-stressed in their load free configuration, and neglect aspects such as fiber-fiber and fiber-matrix cross-links \citep[\eg][]{aver&bail08inbook}.
Nonetheless, adaptation of this approach allows to reflect features of the arteries geometry and morphology,
in particular the orientations distinguished by the collagen fiber structure and variations in their content. 
Among other aspects this approach will also allow to account for changes in the composition of the constituents, thickening of the fibers, alternation in their density, deterioration of the EFM due to disease \citep[\eg][]{defi&etal08hupa} or variations in nutrition habits.
Indeed, the advantages of being able to extract the overall behavior of the tissue from the individual behaviors of the phases and their histology are clear. 
In particular, if fair estimates for the fiber and the extra-fibrillar matrix behaviors are known, this technique will play a key role in providing exact and realistic predictions for the mechanical properties of both healthy and diseased tissues that can be deduced from patient-specific non-invasive inspection techniques. 

Consistent micromechanics analyses are commonly based on homogenization theories.
While numerous works adopting this approach exist in the limit of infinitesimal deformation elasticity, for finitely deforming materials only a few works can be found in the literature.
The foundations of the homogenization theory for geometrically nonlinear materials were outlined 40 years ago by \cite{hill72prsl} and \cite{ogden74jmps}, however, analytical results for fibrous materials that are based on this theory become available only in the last decade.
Rigorous predictions and estimates for hyperelastic materials with one family of fibers were introduced by \cite{gdb&har06plet,gdb&etal06jmps,shmu&gdb10jem} and \cite{lope&idia10jem}, and corresponding results for materials with two families of fibers were determined by \cite{gdb&shmu09jmps}.
Estimates for fiber composites that are based on the linear comparison method of \cite{ppc&tibe00jmps} were determined by \citet{brun&etal07ijss} and \citet{agor&etal09ajmps}.
\cite{chen&etal11jmps} followed this method to estimate the behavior of materials with two fiber families.
The results presented in this work are based on the \emph{nonlinear comparison} (NLC) variational estimate introduced recently by \citet{gdb&shmu10jmps}.
This method allows to extend available estimates for the macroscopic behavior of \emph{hyperleastic} heterogeneous materials to families of hyperelastic materials with different constitutive behaviors of the phases.

To demonstrate the validity of the estimates obtained by application of the NLC variational method we compare its predictions with corresponding finite element (FE) simulations.
In this regard we note that due to the expected advantage of the micromechanics-motivated approach, the microsphere formulation, which is a finite element based technique that accounts for the local microstructure of the material, was recently applied to model the response and remodeling in soft collagenous tissues \citep[\emph{e.g.},][]{menz&waff09ptrsa,alas&etal09jmps}.
Here, we numerically tackle the homogenization problem by considering a 3-D periodic specimen whose representative unit cell contains two families of fibers.
We note that this approach is inefficient for large models, but in this work it allows to compare the predictions of the analytical estimates under uniform loading conditions.

The work is structured as follows: In the following section we briefly revisit the required background including homogenization theory in finite deformation elasticity. 
We also recall the variational statement that enables to determine the analytical estimates for the behavior of the collagenous tissue as well as the complementary finite element procedure for periodic media.
In section 3 we apply the two techniques to a specific tissue, namely a healthy human coronary artery.
This is accomplished by appropriate choices of the behaviors of the constituents composing the artery and its histology.
Finally, predictions for the response of the tissue to different loadings according to the analytical and the numerical formulations are determined and compared.
A short summary concludes the work.

\section{Theoretical Background}\label{sTheoBack}

The Cartesian position vector of a material point in a reference
configuration of a body ${\mathcal{B}_{0}}$ is $\refposT$, and its position vector in
the deformed configuration ${\mathcal{B}}$ is $\curposT$. The deformation of the
body is characterized by the mapping
\begin{equation}\label{chi}
    \curposT=\mapping(\refposT).
\end{equation}
The deformation gradient is
\begin{equation}\label{deformation_gradient_tensor1}
    \defgT=\frac{\partial \mapping(\refposT)}{\partial \refposT}.
\end{equation}
We assume that the deformation is invertible and hence $\defgT$ is
non-singular, accordingly
\begin{equation}\label{J determinant gradF}
    J\equiv\det \defgT \neq 0.
\end{equation}
Physically, $J$ is the volume ratio between the volumes of an
element in the deformed and the reference configurations, and
hence $J=\frac{dv}{dV}>0$.

Soft biological tissues are commonly treated as hyperelastic materials for which the constitutive relations are given in terms of a scalar-valued \emph{strain energy-density function} $\sedf(\defgT)$ such that
\begineq{definition strain energy func}
    \nominalT = \dfrac{\partial\sedf(\defgT)}{\partial\defgT},
\end{equation}
where $\nominalT$ is the first Piola-Kirchhoff or the nominal stress tensor. 
The true or Cauchy stress tensor is related to the first Piola-Kirchhoff stress tensor via the
``push forward'' operation
\begin{equation}\label{Piola-Cauchy}
    \cauchyT=J^{-1}\nominalT\defgT^{T}.
\end{equation}
In the absence of body forces the equilibrium equation is
\begin{equation}\label{equi1}
    \nabla\cdot\cauchyT=0.
\end{equation}

The strain energy-density function of an isotropic material can be expressed in terms of the three invariants of the Cauchy-Green strain tensor $\CGstrain\equiv\defgT^{T}\defgT$, namely
\begin{equation}\label{invariants}
    I_{1}=\Trace \CGstrain,\quad I_{2}=\dfrac{1}{2}(I_{1}^{2}-\Trace
    (\CGstrain^{2})),\quad I_{3}=\det \CGstrain .
\end{equation}
The strain energy-density function of a transversely isotropic (TI) material, whose preferred
direction in the reference configuration is along the unit vector $\refnormalT_{1}$, depends on the two additional
invariants
\begin{equation}\label{addInvar}
    I_{4}=\refnormalT_{1}\cdot\CGstrain\refnormalT_{1}\quad\text{and}\quad I_{5}=\refnormalT_{1}\cdot\CGstrain^{2}\refnormalT_{1}.
\end{equation}
A tissue with two identical families of fibers admits an orthotropic symmetry.
If the two families of fibers are aligned along the unit vectors $\refnormalT_{1}$ and  $\refnormalT_{2}$, the strain energy-density function describing its behavior depends on three more invariants.
Two are analogous to the ones listed in \Eq{addInvar}, namely
\begin{equation}\label{evenmoreInvar}
    I_{6}=\refnormalT_{2}\cdot\CGstrain\refnormalT_{2}\quad\text{and}\quad I_{7}=\refnormalT_{2}\cdot\CGstrain^{2}\refnormalT_{2},
\end{equation}
and the last one
\begin{equation}\label{lastInvar}
    I_{8}=\refnormalT_{1}\cdot\CGstrain\refnormalT_{2},
\end{equation}
 corresponds to the interaction between the two families.

As was mentioned in the introduction biological tissues are heterogeneous materials.
Accordingly, we consider next a heterogeneous body made out of $N$ distinct phases whose behaviors are characterized by the strain energy-density functions $\sedf^{(r)},\ r=1,\ldots,N$.
In the reference configuration each phase occupies a domain ${\mathcal{B}_{0}}^{(r)}$.
The \emph{local} behavior of the heterogeneous material is given in terms of the strain energy-density function
\begin{equation}\label{totalW}
    \sedf(\defgT,\refposT)=\underset{r=1}{\overset{n}\sum}\varphi^{(r)}(\refposT)\sedf^{(r)}(\defgT),
\end{equation}
where the characteristic functions
\begin{equation}\label{vaph}
    \varphi^{(r)}(\refposT)=\left\{%
\begin{array}{ll}
    1\quad \text{if}\ \refposT\in{\mathcal{B}_{0}}^{(r)}, \\
    0\quad\text{otherwise},  \\
\end{array}%
\right.
\end{equation}
describe the spatial distributions of the phases within the body.
The \emph{volume fraction} of the $r$-phase in the material is
\begin{equation}\label{cr}
    \volfrac^{(r)}=\frac{1}{V}\underset{{\mathcal{B}_{0}}}\int \varphi^{(r)}(\refposT)dV,
\end{equation}
where $V$ is the referential volume of the body and clearly $\sum_{r=1}^{N}\volfrac^{(r)}=1$.

For accurate modeling of this heterogeneous material behavior we need, at one hand to account for the interactions between the phases and, at the other, to characterize its macroscopic behavior in terms of relevant information which is elevated from the microscopic level.
To accomplish this task we recall the homogenization procedure introduced in the pioneering works of \citet{hill72prsl} and \citet{ogden74jmps}.
Thus, we apply \emph{homogeneous} boundary conditions 
\begineq{homogeneous bc}
\curposT=\defgT_{0}\refposT
\end{equation}
on the boundary of the heterogeneous body $\partial{\mathcal{B}_{0}}$, where $\defgT_{0}$ is a constant matrix with $\det\defgT_{0}>0$.
By application of the divergence theorem it can be shown that the average deformation gradient satisfies the relation 
\begineq{avgF}
    \bar{\defgT}\equiv\dfrac{1}{V}\underset{{\mathcal{B}_{0}}}{\int}\defgT(\refposT)dV=\defgT_{0}.
\end{equation}
The average nominal stress tensor is
\begin{equation}\label{avgPf}
    \bar{\nominalT}=\dfrac{1}{V}\underset{{\mathcal{B}_{0}}}{\int}\nominalT(\refposT)dV.
\end{equation}
By application of the principle of minimum energy, the \emph{effective} or \emph{macroscopic} strain energy-density function is \citep{hill72prsl,ogden74jmps},
\begineq{Wmin}
    \tilde\sedf(\bar{\defgT})=\underset{\defgT\in \mathcal{K}(\bar{\defgT})}\inf\left\{\dfrac{1}{V}\underset{{\mathcal{B}_{0}}}{\int}\sedf(\defgT,\,
    \refposT)dV\right\},
\end{equation}
where 
$\mathcal{K}(\defgT)\equiv\Bigl\{\defgT\mid\defgT=\frac{\partial\mapping(\refposT)}{\partial\refposT},\,\refposT
\in {\mathcal{B}_{0}};\, \mapping(\refposT)=\bar{\defgT}\refposT,\, \refposT\in\partial{\mathcal{B}_{0}}\Bigr\}$
is the set of kinematically admissible deformation gradients. The
corresponding macroscopic constitutive relation is
\begineq{constMacro}
    \bar{\nominalT} = \dfrac{\partial\tilde\sedf(\bar{\defgT})}{\partial\bar{\defgT}}.
\end{equation}
The associated macroscopic true stress is determined by the ``push forward'' operation \Eq{Piola-Cauchy} with respect to $\bar\defgT$.
Generally, application of the variational principle \eqref{Wmin} to heterogeneous materials can lead to bifurcations corresponding to abrupt changes in the nature of the solution to the optimization problem \citep[\eg][]{tria&make85jamt,rudy&gdb11jelas}.
In this work the issue of instabilities is not accounted for and the response of the heterogeneous media is considered only along the primary loading curve.

The solution to the variational statement \Eq{Wmin} is not trivial, and we will make use of the NLC variational estimate of \citet{gdb&shmu10jmps} to obtain \emph{close form} estimates for the effective behavior of the tissue. 
The NLC variational estimate for $\tilde\sedf (\bar\defgT) $ states that
 \begin{equation}\label{VatEst}
    \hat\sedf(\bar\defgT)=\hat\sedf_{0}(\bar\defgT)+\underset{r=1}{\overset{n}\sum}\volfrac^{(r)}\underset{\defgT}\inf\left\{\sedf^{(r)}(\defgT)-\sedf^{(r)}_{0}(\defgT)\right\},
\end{equation}
where $\hat\sedf_{0}$ is an estimate for the effective SEDF of a comparison hyperelastic heterogeneous material whose phases behaviors are governed by the SEDFs $\sedf^{(r)}_{0}$, and their distributions are characterized by the functions $\varphi^{(r)}$ given in \Eq{vaph}.
The term appearing in the second part of \Eq{VatEst} is denoted the \emph{corrector term}, and we note that it depends only on the properties of the phases composing the two heterogeneous materials.

This method was applied by \citet{gdb&shmu10jmps} and \cite{rudy&gdb11jelas} to determine the behaviors of transversely isotropic materials with one family of fibers.
Thus, it is assumed that there are two distinct constituents in the tissue, the fibrous phase and the extra-fibrillar matrix. 
Each phase is incompressible, isotropic, and its behavior can be approximated by a hyperelastic model that depends on $I_{1}$ only.
Specifically, we denote by $\sedf^{(f)}(I_{1})$ and $\sedf^{(m)}(I_{1})$ the strain energy-density functions of the fiber and the extra-fibrillar matrix phases, respectively.
The volume fractions of the two phases are $\volfrac^{(f)}$ and $\volfrac^{(m)}=1-\volfrac^{(f)}$.
As the comparison media the model for neo-Hookean fibrous materials introduced by \citet{gdb&etal06jmps} is employed \citep[see also][]{gdb05jmps,lope&idia10jem}.
Following \cite{gdb&shmu10jmps}, the NLC estimate for the effective strain energy-density function is given in terms the minimization problem
\begin{equation}\label{fiber material}
    \hat\sedf^{(TI)}\left(\bar\defgT\right)=\min_{\omega_{1}}\left\{\volfrac^{(m)}\sedf^{(m)}\left({\tilde{I}_{1}^{(m)}}\left(\BI{1},\BI{4},\omega_{1}\right)\right)+\volfrac^{(f)}\sedf^{(f)}\left({\tilde{I}_{1}^{(f)}}\left(\BI{1},\BI{4},\omega_{1}\right)\right)\right\},
\end{equation}
where 
\begineq{I11}
    {\tilde{I}_{1}^{(r)}}\left(\BI{1},\BI{4},\omega_{1}\right)=
    \BI{4}+\frac{2}{\sqrt{\BI{4}}}+{\alpha}^{(r)}(\omega_{1})\left({\BI{1}-\BI{4}-\frac{2}{\sqrt{\BI{4}}}}\right),
\end{equation}
$\BI{1}$ and $\BI{4}$ are the invariants of the macroscopic Cauchy-Green strain tensor $\bar\CGstrain=\bar\defgT^{T}\bar\defgT$, and
\begineq{alphaF}
    {\alpha}^{(f)}(\omega)=\left(1-\volfrac^{(m)}\omega\right)^{2}
    \quad \mathrm{and}\quad
    {\alpha}^{(m)}(\omega)=\left(1+\volfrac^{(f)}\omega\right)^{2}+\volfrac^{(f)}\omega^{2}.
\end{equation}
Denote by $\tilde\omega_{1}$ the value of $\omega_{1}$ that yields the solution to the minimization problem \eqref{fiber material}.
In some cases $\tilde\omega_{1}$ can be determined analytically otherwise it must be calculated numerically. 
In general, however, $\tilde\omega_{1}=\tilde\omega_{1}(\bar\defgT)$.
Nonetheless, since the partial derivative of $\hat\sedf^{(TI)}$ with respect to $\omega_{1}$ identically vanishes at $\tilde\omega_{1}$, the expression for the macroscopic stress can be evaluated \emph{analytically} without the need for differentiating $\tilde\omega_{1}$ with respect to $\bar\defgT$.
Consequently, the macroscopic nominal stress is
\begineq{stressP}
\begin{split}
    {\nominalT}^{(TI)}=&
    \underset{r=m,\,f}\sum 2\volfrac^{(r)}
  \sedf_{1}^{(r)}\left({\tilde{I}_{1}^{(r)}\left(\BI{1},\BI{4},\tilde\omega_{1}\right)}\right)
  \left[{{\alpha}^{(r)}(\tilde\omega_{1})} \bar\defgT+\left(1-{{\alpha}^{(r)}(\tilde\omega_{1})}\right)\left(1-{\BI{4}}^{\ -\frac{3}{2}}\right)\bar\defgT\refnormalT\otimes\refnormalT\right]\\
    &+p\bar\defgT^{-T},
\end{split}
\end{equation}
where $\sedf_{1}^{(r)}\left({I_{1}}\right)$ is the derivative of $\sedf^{(r)}\left({I_{1}}\right)$ with respect to its argument.

\begin{figure}
\begin{center}
 (a) \includegraphics[scale=0.245]{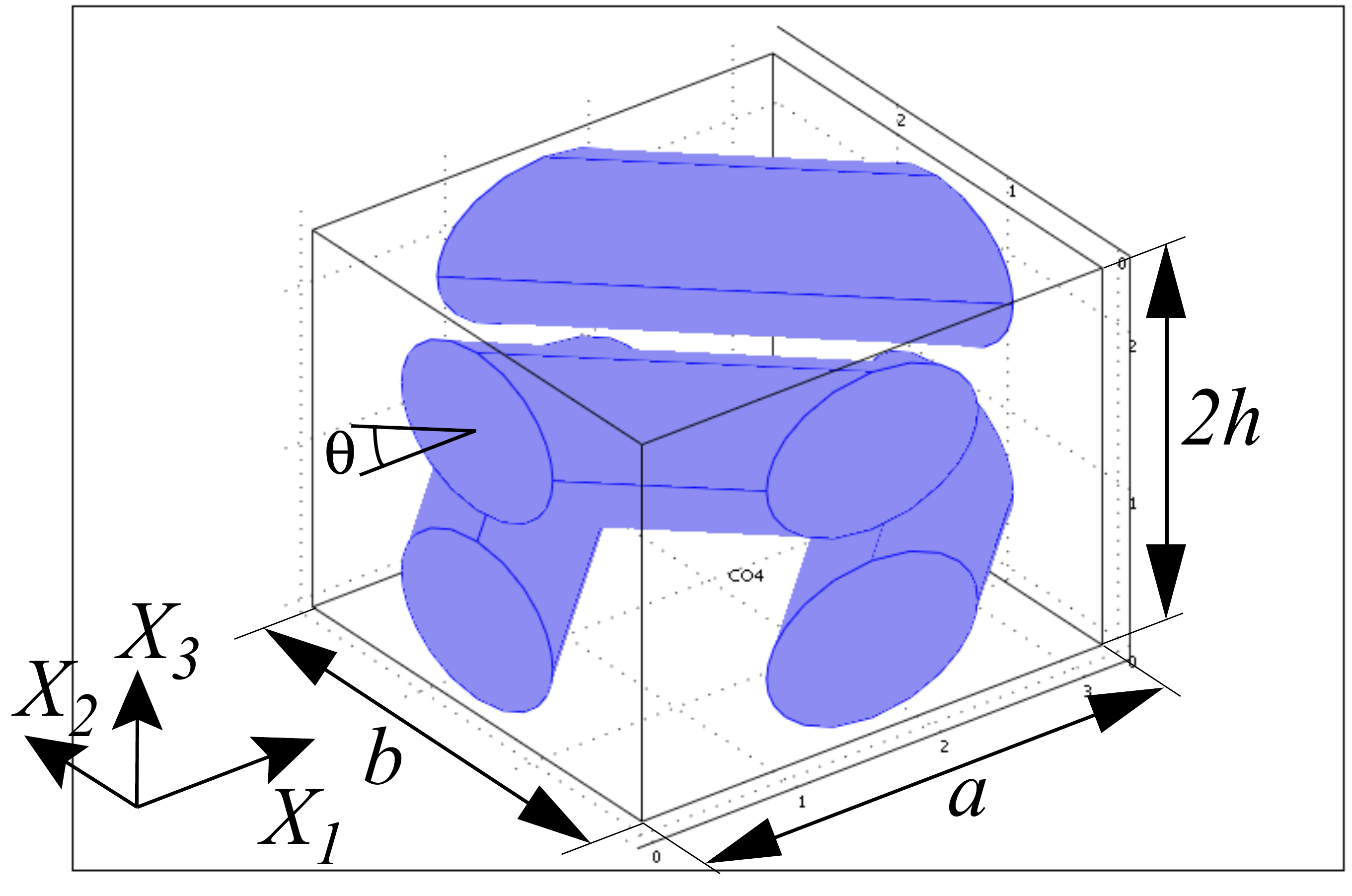}
 (b) \includegraphics[scale=0.25]{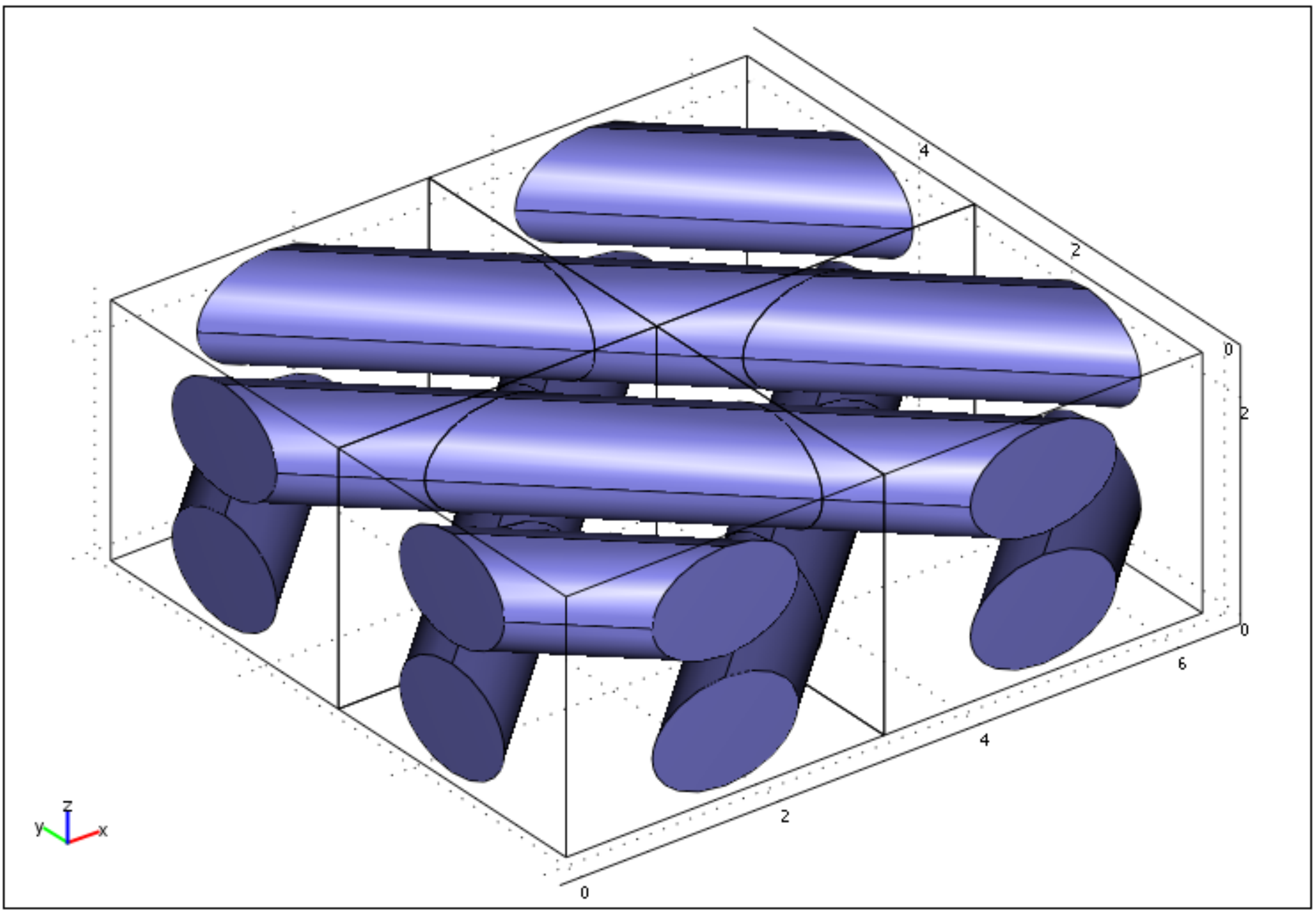}\\
  \caption{\footnotesize Schematic drawings of (a) the unit cell and (b) a section of the periodic media with four unit cells.
}\label{UC}
  \end{center}
\end{figure}

We model a tissue with two families of fibers as a layered medium made out of transversely isotropic fibrous layers such that in the referential configuration the fibers in two adjacent layers are alternately directed along the vectors $\refnormalT_{1}$ and $\refnormalT_{2}$ \citep[\eg][]{gdb&shmu09jmps}. 
We stress that in contrast with artificial composites where, due to the manufacturing process, the interaction is between the two layers of fibers, in biological tissues the interaction is between the individual fibers belonging to the two families (\eg~Fig.~\ref{UC}b).
One of the aims of this work is to investigate the role of these different interaction hierarchies and the effect of this assumption on the predicted overall response.
It is assumed that the thickness of the layers is substantially smaller than the thickness of the tissue, and that the layers are perfectly bonded to each other.
In a tissue with two identical families of fibers the unit vectors are symmetrically aligned with respect to a symmetry axis (\eg~$X_{1}$-axis in Fig.~\ref{UC}a), and the volume fractions of the two layers are identical.
The overall symmetry of this two-hierarchies model is orthotropic.

In a physiological state it is anticipated that the anti-plane shear components are orders of magnitude smaller than the in-plane components (\eg~in the coordinate system of Fig.~\ref{UC}a, $\defg_{13}$ and $\defg_{23}$ are substantially smaller than the other components of $\defgT$).
Accordingly, as outlined in Appendix~\ref{laminate}, the overall behavior of the orthotropic tissue is
\begineq{orthotropic sedf}
\begin{split}
    \hat\sedf^{(TFF)}\left(\bar\defgT\right)=&
    \frac{1}{2}\min_{\omega_{1}}\left\{\volfrac^{(m)}\sedf^{(m)}\left({\tilde{I}_{1}^{(m)}}\left(\BI{1},\BI{4},\omega_{1}\right)\right)+\volfrac^{(f)}\sedf^{(f)}\left({\tilde{I}_{1}^{(f)}}\left(\BI{1},\BI{4},\omega_{1}\right)\right)\right\}+\\
    &\frac{1}{2}\min_{\omega_{2}}\left\{\volfrac^{(m)}\sedf^{(m)}\left({\tilde{I}_{1}^{(m)}}\left(\BI{1},\BI{6},\omega_{2}\right)\right)+\volfrac^{(f)}\sedf^{(f)}\left({\tilde{I}_{1}^{(f)}}\left(\BI{1},\BI{6},\omega_{2}\right)\right)\right\},
\end{split}
\end{equation}
where $\BI{1}$, $\BI{4}$ and $\BI{6}$ are the appropriate orthotropic invariants of $\bar\CGstrain=\bar\defgT^{T}\bar\defgT$.
Here, ${\tilde{I}_{1}^{(r)}}\left(\BI{1},\BI{4},\omega_{1}\right)$ and ${\tilde{I}_{1}^{(r)}}\left(\BI{1},\BI{6},\omega_{2}\right)$ are given in \Eq{I11} with $\BI{6}$ and $\omega_{2}$ replacing $\BI{4}$ and $\omega_{1}$ for the second family of  fibers, respectively. 

We note that the two minimization problems in the expression for $\hat\sedf^{(TFF)}$ can be solved independently, and denote by $\tilde\omega_{1}$ and $\tilde\omega_{2}$ the values of $\omega_{1}$ and $\omega_{2}$ at the minima, respectively.
Thanks to an argument similar to the one preceding \Eq{stressP}, the following expression for the macroscopic nominal stress is derived analytically.
\begineq{stressTFF}
\begin{split}
    \hat{\nominalT}^{(TFF)}&=
    \underset{r=m,\,f}\sum \volfrac^{(r)}
  \sedf_{1}^{(r)}\left({\tilde{I}_{1}^{(r)}\left(\BI{1},\BI{4},\tilde\omega_{1}\right)}\right)
  \left[{{\alpha}^{(r)}(\tilde\omega_{1})} \bar\defgT+\left(1-{{\alpha}^{(r)}(\tilde\omega_{1})}\right)\left(1-{\BI{4}}^{\ -\frac{3}{2}}\right)\bar\defgT\refnormalT_{1}\otimes\refnormalT_{1}\right]\\
   &+\underset{r=m,\,f}\sum \volfrac^{(r)}
  \sedf_{1}^{(r)}\left({\tilde{I}_{1}^{(r)}\left(\BI{1},\BI{6},\tilde\omega_{2}\right)}\right)
  \left[{{\alpha}^{(r)}(\tilde\omega_{2})} \bar\defgT+\left(1-{{\alpha}^{(r)}(\tilde\omega_{2})}\right)\left(1-{\BI{6}}^{\ -\frac{3}{2}}\right)\bar\defgT\refnormalT_{2}\otimes\refnormalT_{2}\right]\\
    &+p\bar\defgT^{-T},
\end{split}
\end{equation}
where, as before, $\sedf_{1}^{(r)}\left({I_{1}}\right)$ is the derivative of $\sedf^{(r)}\left({I_{1}}\right)$ with respect to its argument.
We emphasize that this is a close-form explicit expression for the stress and the values for $\tilde\omega_{1}$ and $\tilde\omega_{2}$ can be substituted once the minimization problem \Eq{orthotropic sedf} is solved, either analytically or numerically.

Before we proceed we emphasize that the expressions for the macroscopic SEDF $\hat\sedf^{(TFF)}$ and the macroscopic stress $\hat{\nominalT}^{(TFF)}$ involve, as they should, the histology of the tissue in terms of the direction and volume fraction of the collagen fibers together with the expressions for the behaviors of the two constituents in terms of their SEDFs $\sedf^{(m)}$ and $\sedf^{(f)}$.

We consider next the FE model to which the aforementioned analytical expressions are compared.
Drawings of the elementary unit cell and a representative section of the periodic media containing four elementary cells together with the chosen coordinate system are shown in Fig.~\ref{UC}a and b, respectively.
Arbitrarily we define a \emph{material} coordinate system such that in the reference state the direction
between the two families is aligned with the $X_{1}$-axis.
With respect to an artery, axes $X_{1}$ and $X_{2}$ may be associated with the circumferential and the axial directions, respectively.

The referential angle between the fiber's direction and the $X_{1}$-axis is denoted $\Theta$. 
Accordingly, the ratio between the lengths of the planar edges of the cell is
\begin{equation}
\frac{b}{a}=\tan{\Theta},
\end{equation}
where $a$ and $b$ are the lengths of the faces along the $X_{1}$ and the $X_{2}$ axes, respectively.
Due to the periodicity, the total length of the fiber segments in the unit cell is $2\sqrt{a^{2}+b^{2}}$, and the total volume of the fiber in the cell is 
\begin{equation}
V_{f}=\frac{\pi}{2}{d^{2}}\sqrt{a^{2}+b^{2}},
\end{equation}
where $d$ is the diameter of the fibers.
Once the volume fraction of the fiber $\volfrac^{(f)}$ is set, the condition for the distance between the planes of the two families $h$ (half the hight of the unit cell) is
\begin{equation}
\left(\frac{h}{d}\right)=\frac{\pi}{4 \volfrac^{(f)} \sin\Theta}\left(\frac{d}{a}\right).
\end{equation}
The geometrical requirements $h/d>1$ and $a \sin\Theta/d>1$ imply that
\begineq{geometrical requirements}
1<\frac{h}{d}<\frac{\pi}{4 \volfrac^{(f)}}.
\end{equation}
We note that in the $(X_{1},X_{2})$-plane the spacing between the fibers is $a \sin\Theta$, and the spacing between the fibers in the $X_{3}$ direction is $h$.
A requirement that the spacing will be identical leads to
\begineq{identical spacing}
\frac{h}{d}=\sqrt{\frac{\pi}{4 \volfrac^{(f)}}}.
\end{equation}

The response of the heterogeneous material is obtained by applying periodic boundary conditions on a single unit cell (\eg~ Fig.~\ref{UC}a).
In the undeformed configuration the unit cell occupies the domain 
\begin{equation}
0\leqslant X_{1} \leqslant a,\quad 0\leqslant X_{2}\leqslant b,\quad 0\leqslant X_{3}\leqslant 2h.
\end{equation}
The expressions for these conditions are written in terms of the components of the macroscopic deformation
gradient tensor $\bar{\defgT}$ as follows:
\begin{enumerate}
\item The displacements of points on the right ($X_{1}=a$) and the left ($X_{1}=0$) faces are related via 
\begin{equation}\label{PBCr}
\begin{cases}
u_{1}^L=u_{1}^{R}+\Bigl(\bar\defg_{11}-1 \Bigr)a \\
u_{2}^L=u_{2}^{R}+\bar\defg_{21}a \\
\end{cases},
\end{equation}
and traction free conditions were assumed for the shear components in the $X_{3}$ direction.
\item The displacements of points on the front ($X_{2}=0$) and the rear ($X_{2}=b$) faces are related via 
\begin{equation}\label{PBCf}
\begin{cases}
u_{1}^F=u_{1}^{RE}+\bar\defg_{12}b \\
u_{2}^F=u_{2}^{RE}+  \Bigl(\bar\defg_{22}-1 \Bigr)b\\
\end{cases},
\end{equation}
and traction free conditions were assumed for the shear components in the $X_{3}$ direction.
\item On the top ($X_{3}=2h$) and bottom ($X_{3}=0$) faces $u_{3}^{B}=u_{3}^{T}=0$, and traction free conditions were assumed for the shear components in the $X_{1}$ and $X_{2}$ directions.
\end{enumerate}
The overall behavior of the heterogeneous media was then determined from the relations between the average of the stress field (\emph{e.g.}, \refeq{avgPf} with \refeq{Piola-Cauchy}), and the corresponding applied macroscopic deformation $\bar\defgT$.

For inclined loadings we distinguish, for convenience, between the \emph{principal} coordinate system of the right Cauchy-Green deformation tensor and the material coordinate system.
Under plane-strain loading condition, thanks to the assumed incompressibility of the tissue, in the principal coordinate system the most general macroscopic deformation gradient is
\begineq{fprime}
\bar\defgT'=\left(\begin{array}{ccc}
\lambda & 0 & 0\\
0 & {\lambda}^{-1} & 0\\
0 & 0 & 1\end{array}\right),
\end{equation}
where $\lambda$ is the principal stretch ratio.
This is related to the average deformation gradient in the material coordinate system via the relation
\begin{equation}\label{F_avg}
    \bar\defgT=\Rotation^{T}{\bar{\defgT}'}\Rotation,
\end{equation}
where
\begin{equation}\label{Qort}
    \Rotation=\left(%
\begin{array}{ccc}
\cos\rotangle & -\sin\rotangle & 0\\
\sin\rotangle & \cos\rotangle & 0\\
0 & 0 & 1
\end{array}%
\right),
\end{equation}
and $\rotangle$ is the referential angle between the direction of the principal stretch ratio and the axis $X_{1}$.

\section{Applications}\label{sAppl}

In this section we make use of both the variational estimate and the periodic FE simulation technique to determine the overall behavior of a specific tissue with two families of fibers.
To this end we first make an appropriate choice for the SEDFs of the collagen fibers and the extra-fibrillar matrix.

While most of the biochemical processes in the tissue occur within the extra-fibrillar matrix, from a mechanical viewpoint the fiber phase is more interesting. 
In the referential state of the tissue the fibers are crimped and straighten when the tissue is subjected to tension in the fiber direction \citep[\eg][]{free&doeh05jbet,holz08book}.
Thus, initially the contribution of the fiber is quite small but when straighten it becomes the primary load carrying component in the tissue.
Since not all the fibers straighten simultaneously, there is a recruitment stage during which more and more fibers straighten and gradually increase the overall stiffness of the tissue.
A hyperelastic model that can capture this phenomenon is the \citet{gent96rc&t} model
\begin{equation}\label{Gent}
    \sedf_{G}{(\defgT;\mu,J_{m})} = -\dfrac{1}{2}\mu J_{m}\ln{\Bigl(1-\dfrac{I_{1}-3}{J_{m}}\Bigr)}.
\end{equation}
Here $\mu$ is a shear like modulus and $J_{m}$ is a dimensionless ``locking'' parameter.
Qualitatively, the behavior exhibited by this model can be characterized as follows.
In the small strains regime when $I_{1}\ll J_{m}+3$, the linear term of Taylor series expansion of \Eq{Gent}, namely $\frac{1}{2}\mu I_{1}$, is the dominating term.
In accordance with the recruitment processes, as $I_{1}$ increases the material stiffens.
Finally, in the limit $(I_{1}-3)\rightarrow J_{m}$ there is a dramatic stiffening and the material locks up as $\sedf_{G}$ becomes unbounded.
For example, in the pure shear loading condition of \Eq{fprime} the critical stretch ratio at which locking occurs is $\lambda_{c}=\frac{1}{2}\left(\sqrt{J_{m}+4}+\sqrt{J_{m}}\right)$.

According to the findings of \citet{gund&etal07jbiomech} a hyperelastic model that can capture the behavior of the extra-fibrillar matrix with the elastin fibers is the incompressible neo-Hookean model 
\begin{equation}\label{nH}
    \sedf_{H}(\defgT;\mu) = \dfrac{\mu}{2}\Bigl(I_{1}-3\Bigr).
\end{equation}
This model is characterized by a single shear-like modulus $\mu$. In the limit $J_{m}\rightarrow\infty$ the Gent model \eqref{Gent} reduces to the neo-Hookean model.

In the course of this work models of tissues with neo-Hookean matrices were analyzed by both the NLC variational estimate and the finite element method, and compared with corresponding predictions of phenomenological models.
It was found that as long as no dispersion of the fibers is assumed the usage of a neo-Hookean model for the matrix phase leads to a fine agreement with the phenomenological predictions.
However, as soon as the distributions of the fibers about their primary directions are accounted for, the neo-Hookean model fails to provide satisfactory agreement with the experimentally fitted findings.
The reason is the inability of the neo-Hookean model to capture the stiffening of the dispersed fibers during extensions that are not along the primary directions.
Consequently, in the current study that is aimed towards demonstration of the ability of the micromechanics based model to capture the essential aspects of the behavior of heterogeneous soft materials, we assume no dispersion of the fiber and a Gent model for the behavior of the extra-fibrillar matrix.
The neo-Hookean model will be used in subsequent studies where dispersion of the fibers will be accounted for.

Aside from the ability of the chosen Gent model to capture the behaviors of the two constituting phases, a computational advantage stems from the fact that with this choice the optimization problem \eqref{orthotropic sedf} can be solved \emph{analytically}.
Thus, the associated Euler-Lagrange equations admit the form of cubic polynomials in $\omega_{1}$ and $\omega_{2}$, from which explicit close-form expressions for $\tilde\omega_{1}$ and $\tilde\omega_{2}$ are obtained \citep{gdb&shmu10jmps}.
We emphasize, however, that within the framework of the proposed variational estimate \eqref{orthotropic sedf}, the models for the tissue constituents can be easily modified or replaced with other, possibly more sophisticated, models.

In total there are 4 material parameters corresponding to the shear moduli and the locking parameters of the extra-fibrillar matrix and the fibers.
In addition, there are two histological parameters, namely the volume fraction of the fibers and the opening angle between the two families of fibers.
The parameters were chosen by fitting the overall prediction of the micromechanics based models to the predictions of a phenomenological model whose own parameters were fitted to experimental data.
Variety of results are available in the literature among which we recall the works of 
\cite{holz06thebio,har&etal07bmmb,kiou&etal09jbta,stal09bmmb} and \cite{mort&etal10abe}.
Here we calibrated the micromechanics model with the aid of the strain energy-density function given in \cite{holz&etal05ajphcp} that accounts for the dispersion of the fibers.
This model involves five materials parameters.
Mean values for these parameters are also given in \cite{holz&etal05ajphcp} for the three layers composing the wall of healthy human coronary arteries.
From a mechanical viewpoint the relevant layers are the Adventitia and the Media \citep{holz08book}.
At this stage of our study we wish to treat the tissue as a single layer and hence we used the average value of the parameters for the two dominant layers.
Thus, the numerical values used for the five parameters appearing in Eq.~(1) of \cite{holz&etal05ajphcp} are $\mu=4.42\,\mathrm{kPa}$, $k_{1}=30.1\,\mathrm{kPa}$, $k_{2}=46.6$, $\rho=0.4$ and half the opening angle between the two families of fibers $\Theta=43^{\circ}$.
Note that these values are also in fair agreement with corresponding results that were reported in \cite{stal09bmmb}.
Additionally, \cite{holz&etal00jela,balz&etal06ijss,holz06thebio} and \cite{har&etal07bmmb} reported quite similar estimates for the angle between the two fiber families.
The estimate for the collagen volume fraction was deduced from the experimental findings of \cite{defi&etal08hupa} and the fitting process of \cite{gdb&shmu09jmps}.

\begin{figure}[t]
\begin{center}
  (a)\includegraphics[scale=0.45]{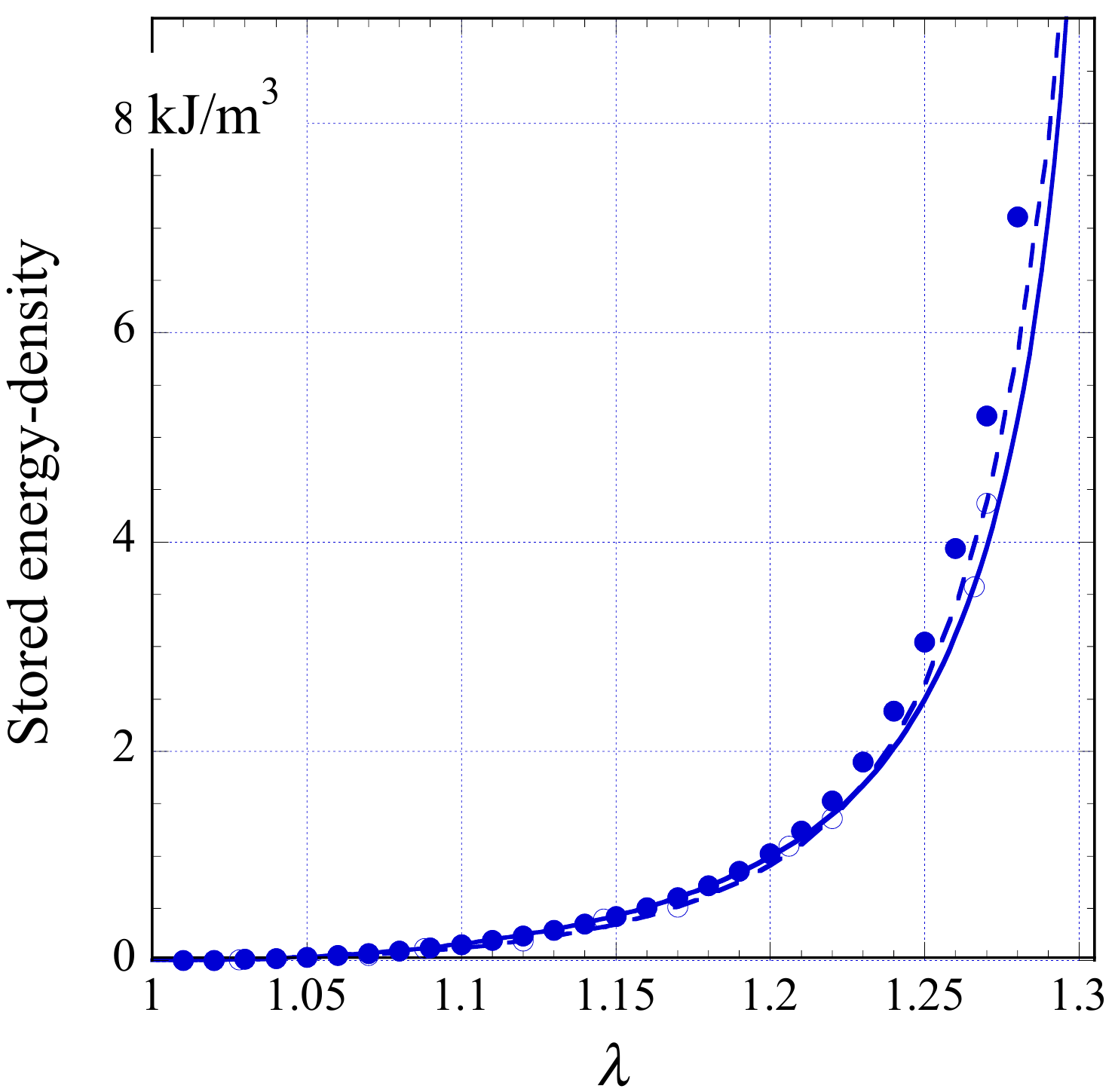}
  \quad
  (b)\includegraphics[scale=0.45]{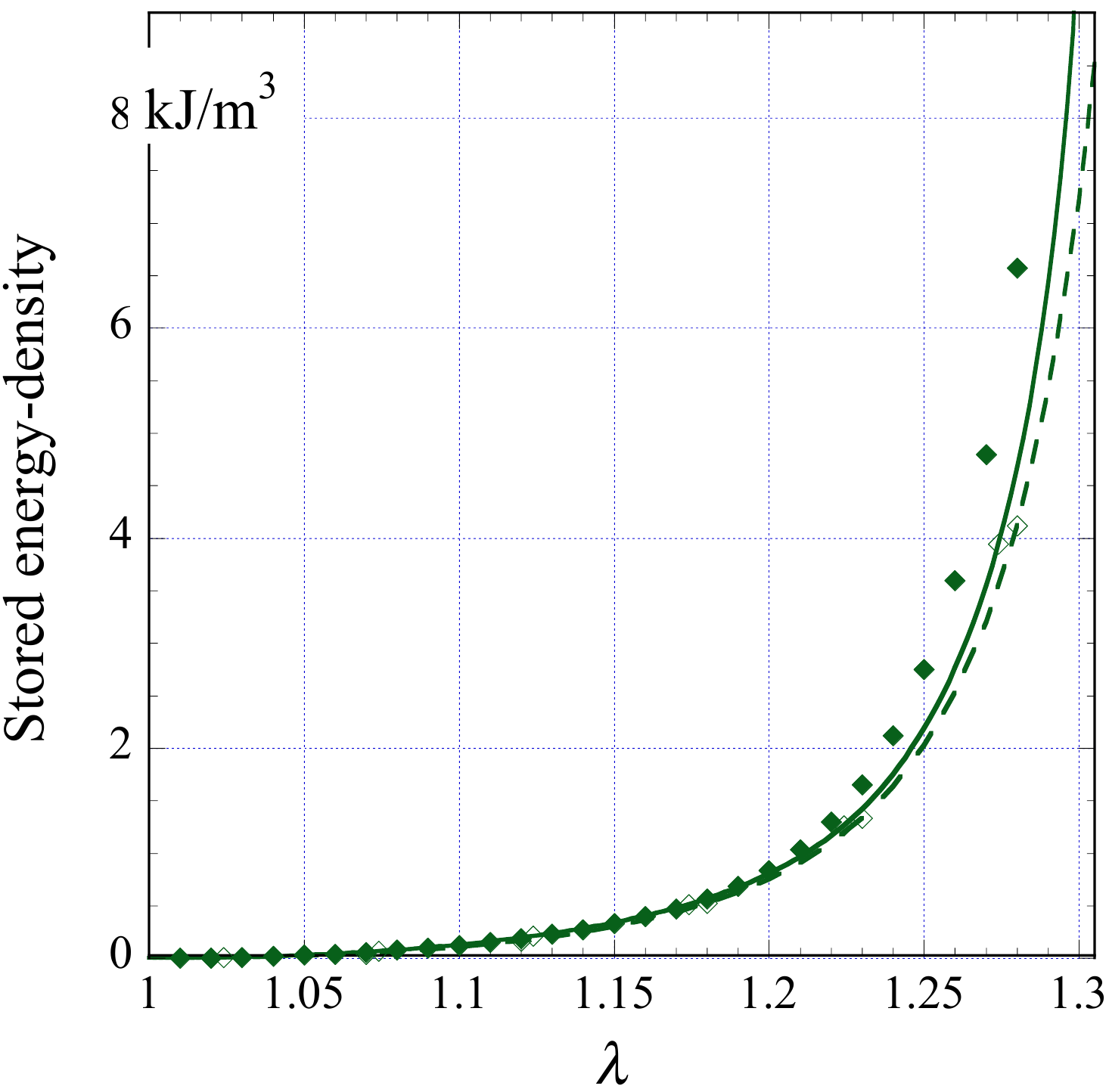}
  \caption{\footnotesize Stored elastic energy-density as a function of the stretch ratio according to NLC estimate (continuous curves), the periodic FE model (shaded marks), and the model of \citet{holz&etal05ajphcp} (dashed curves).
Figures (a) and (b) correspond to extensions along the in-plane principal orthotropic axes $X_{1}$ and $X_{2}$, respectively.}\label{stored_energy_setup_0-90}
\end{center}
\end{figure}
\begin{table}[b]
\caption{Material parameters}
\begin{center}
\begin{tabular}
{|p{0.12\linewidth}|p{0.12\linewidth}|p{0.12\linewidth}|p{0.12\linewidth}|p{0.12\linewidth}|p{0.12\linewidth}|}
\hline 
{ $\mum$} & { $J_{m}^{(m)}$} & { $\muf$} & { $J_{m}^{(f)}$} & { $\volfrac^{(f)}$} & { $\Theta$} \\
\hline 
$3.5$\,kPa & $0.098$ & $280$\,kPa & $0.45$ & $0.30$ & $43^{\circ}$ \\
\hline 
\end{tabular}
\end{center}
\label{parameters_table} 
\end{table}
The numerical values that were chosen for the modeled tissue are summarized in table \ref{parameters_table}.
These were determined by comparing the predictions for the stored energy-density according to the micromechanics based model \Eq{orthotropic sedf} and the FE periodic model with the model of \cite{holz&etal05ajphcp} during extensions along the principal orthotropic axes.
This comparison is depicted in Fig.~\ref{stored_energy_setup_0-90}.
The continuous curves depict the predictions of the NLC estimate, the dark circle and diamond marks those of the periodic FE model, and the dashed curves to the model of \citet{holz&etal05ajphcp}.
Figs.~\ref{stored_energy_setup_0-90}(a) and (b) correspond to extensions along the in-plane principal orthotropic axes $X_{1}$ and $X_{2}$, respectively.
We note that this choice of parameters leads to a fine agreement between the three different models and  emphasize that throughout this work the parameters listed in Table~\ref{parameters_table} are used for both the NLC variational estimate and the FE simulation.
We recall that for the periodic FE model an additional parameter, namely the relative spacing between the fibers, needs to be set.
For the chosen histological parameters inequality \eqref{geometrical requirements} is $1<h/d<2.6$, and according to \refeq{identical spacing} the choice $h/d=1.62$ will lead to identical spacing.
We note that a choice of $h/d<1.62$ will increase the cross-families interaction between the fibers in the two families, whereas a reversed choice will increase the internal interaction between the fibers in the two families.
In this work the choice $h/d=1.2$ was made in order to maximize the interaction between the fibers in the two families.
This was done in order to critically examine the assumption embedded in the analytical model that the out of plane interaction is between layers of fibers and not between the fibers themselves.

\begin{figure}[t]
\begin{center}
  (a)\includegraphics[scale=0.28]{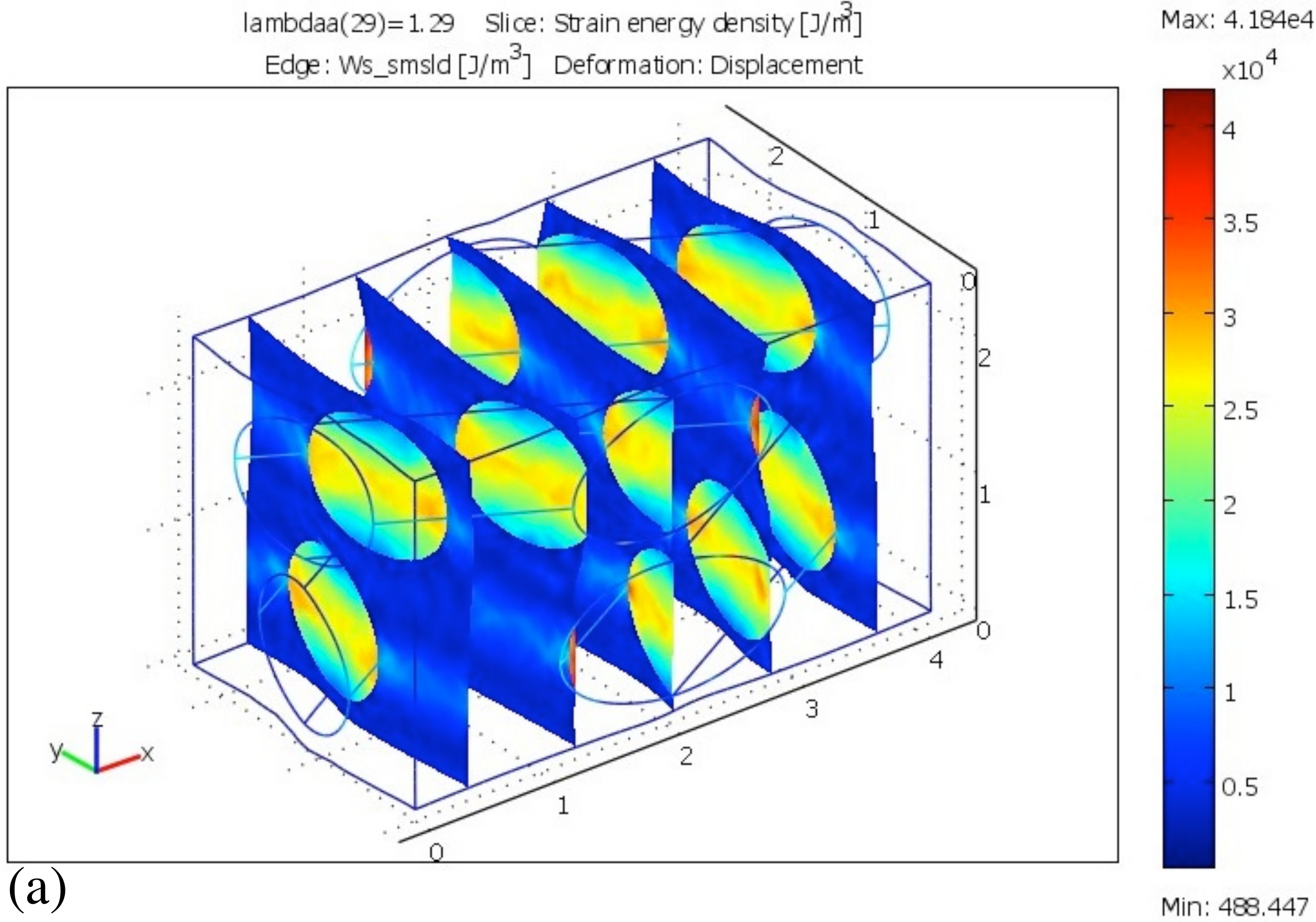}
  \quad
  (b)\includegraphics[scale=0.28]{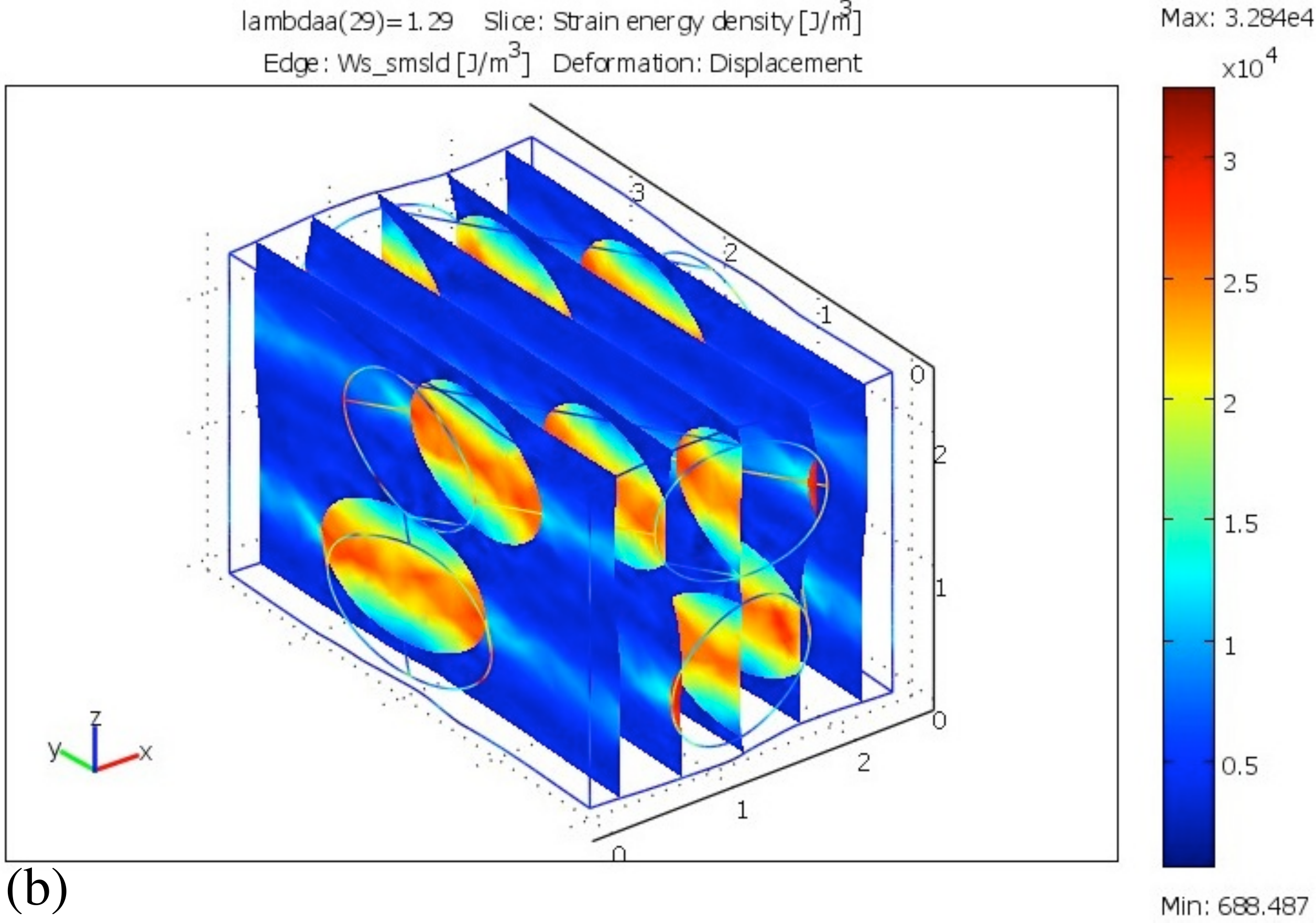}
  \caption{\footnotesize The deformed configuration of the unit cell during extensions alon (a) the $X_{1}$ and (b) the $X_{2}$ directions. The stored energy-density at several slices is depicted by the contour plots.}
  \label{drawing_principal_extensions}
\end{center}
\end{figure}
The deformations of the unit cell and the distributions of the stored strain energy during the two loading conditions are demonstrated in Fig.~\ref{drawing_principal_extensions} for $\lambda=1.29$.
In both cases the fibers are stretched and the elastic energy stored in the fibers is an order of magnitude higher than the one in the EFM.
This is due to the fact that the stiffer fibers carry most of the mechanical load at this stage of the deformation.
It is interesting to note that although the stretching is along the principal orthotropic axes, the cell undergoes local shear and hence in the deformed configuration its boundary is not straight.
This local shear is associated with the imposed periodic boundary conditions.

To examine the ability of the NLC estimate to capture the behavior predicted by the periodic FE simulation, we also compare their predictions under incline loading conditions. 
Two configurations are considered corresponding to tensions at angles $\rotangle=21.5^{\circ}$ and $\rotangle=43^{\circ}$ relative to the principal orthotropic $X_{1}$-axis.
These loading states may correspond to combinations of axial and hoop extensions together with a twist of the artery.
Particularly, with $\rotangle=43^{\circ}$ the tension is along the first fiber families (the bottom family in Fig.~\ref{UC}), while the second family is in a state of compression in the fiber direction.
Under tension at an angle $\rotangle=21.5^{\circ}$, the first family is under a combination of shear and tension while the second family is under compression and shear.

\begin{figure}[t]
\begin{center}
  (\emph{i})\includegraphics[scale=0.28]{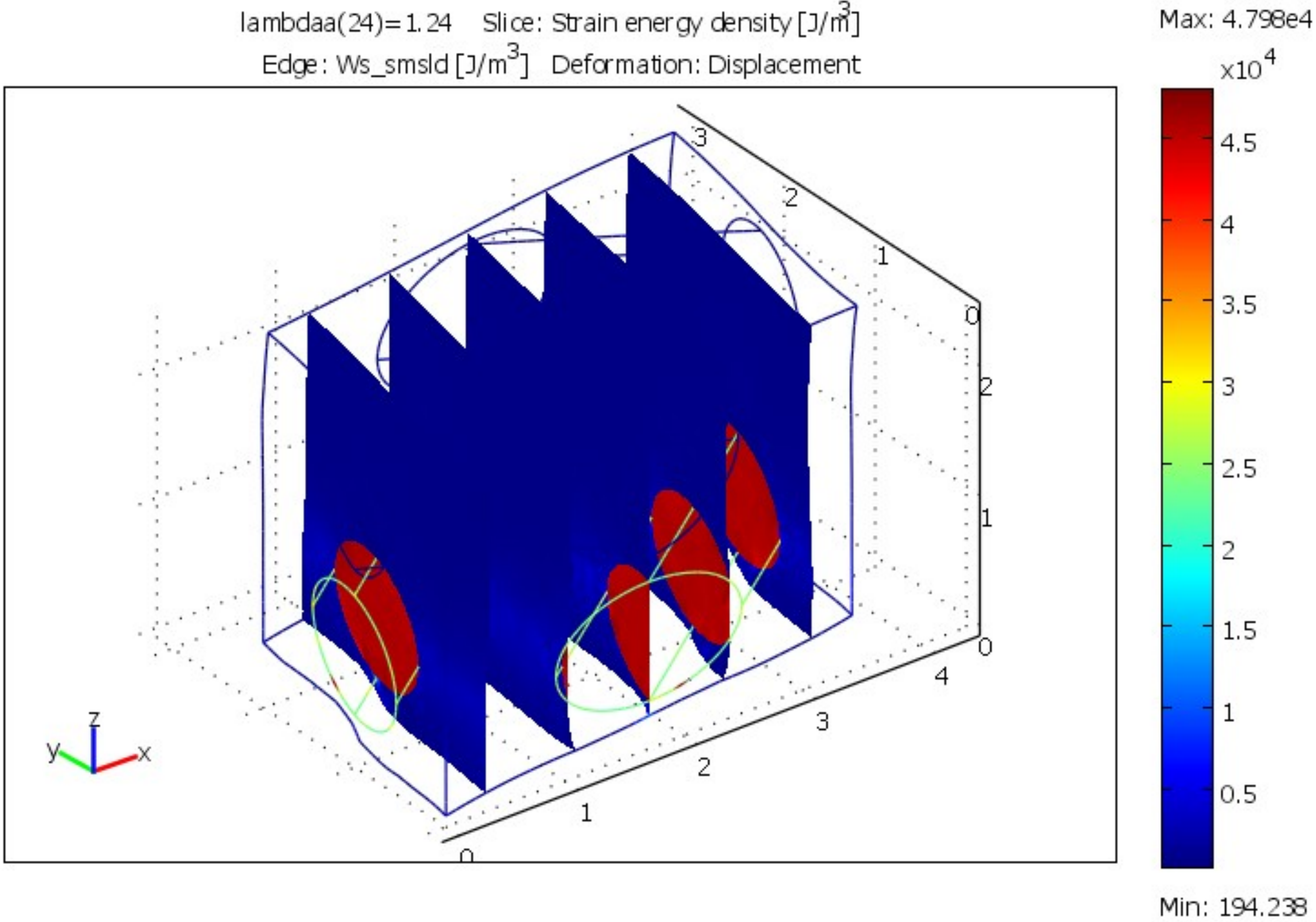}
  \quad
  (\emph{ii})\includegraphics[scale=0.28]{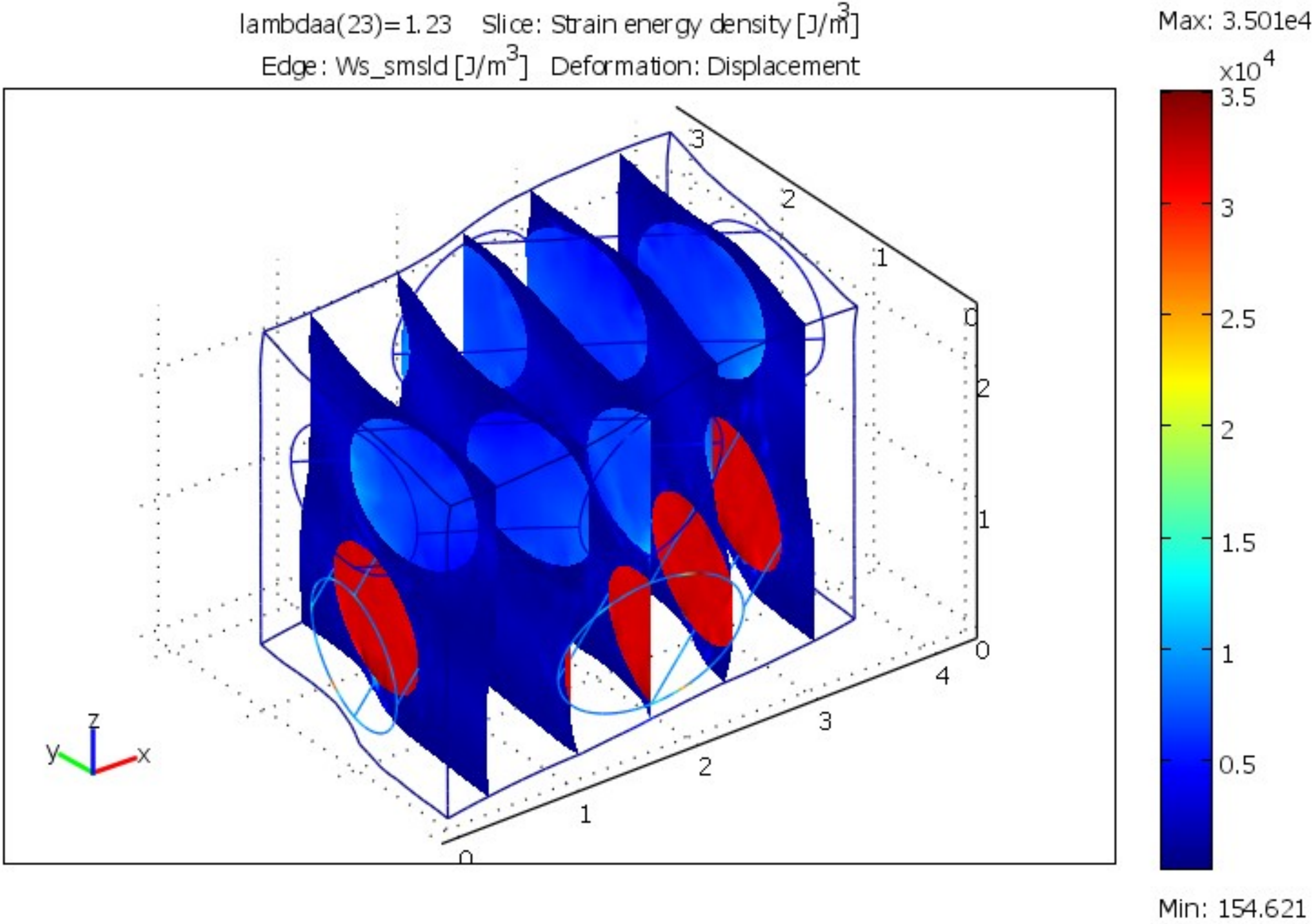}
  \caption{\footnotesize Deformed configuration of the unit cell during incline extension at $\rotangle=21.5^{\circ}$. Case-(\emph{i}) is shown in the left figure and case-(\emph{ii}) in the right one. The stored energy-density at several slices is depicted by the contour plots.}\label{drawing_incline_21}
\end{center}
\end{figure}
It is widely accepted that the collagen fibers do not contribute to the stiffness of the tissue when compressed in the fiber direction \citep[\eg][]{holz08book}.
Accordingly, during the incline tests we examined two different cases.
In case-(\emph{i}) the behavior of the compressed fiber family was set to be identical to that of the extra-fibrillar matrix.
That is, under compression the fiber behaves according to the Gent model with $\mum$ and $J_{m}^{(m)}$.
Essentially, this resulted in models with a single load-carrying family of fibers, the one under tension.
In case-(\emph{ii}) the fiber behavior was not modified and the contribution of the second family to compression was accounted for (\ie~the fiber behaves according to the Gent model with $\muf$ and $J_{m}^{(f)}$).

\begin{figure}[t]
\begin{center}
  (\emph{i})\includegraphics[scale=0.28]{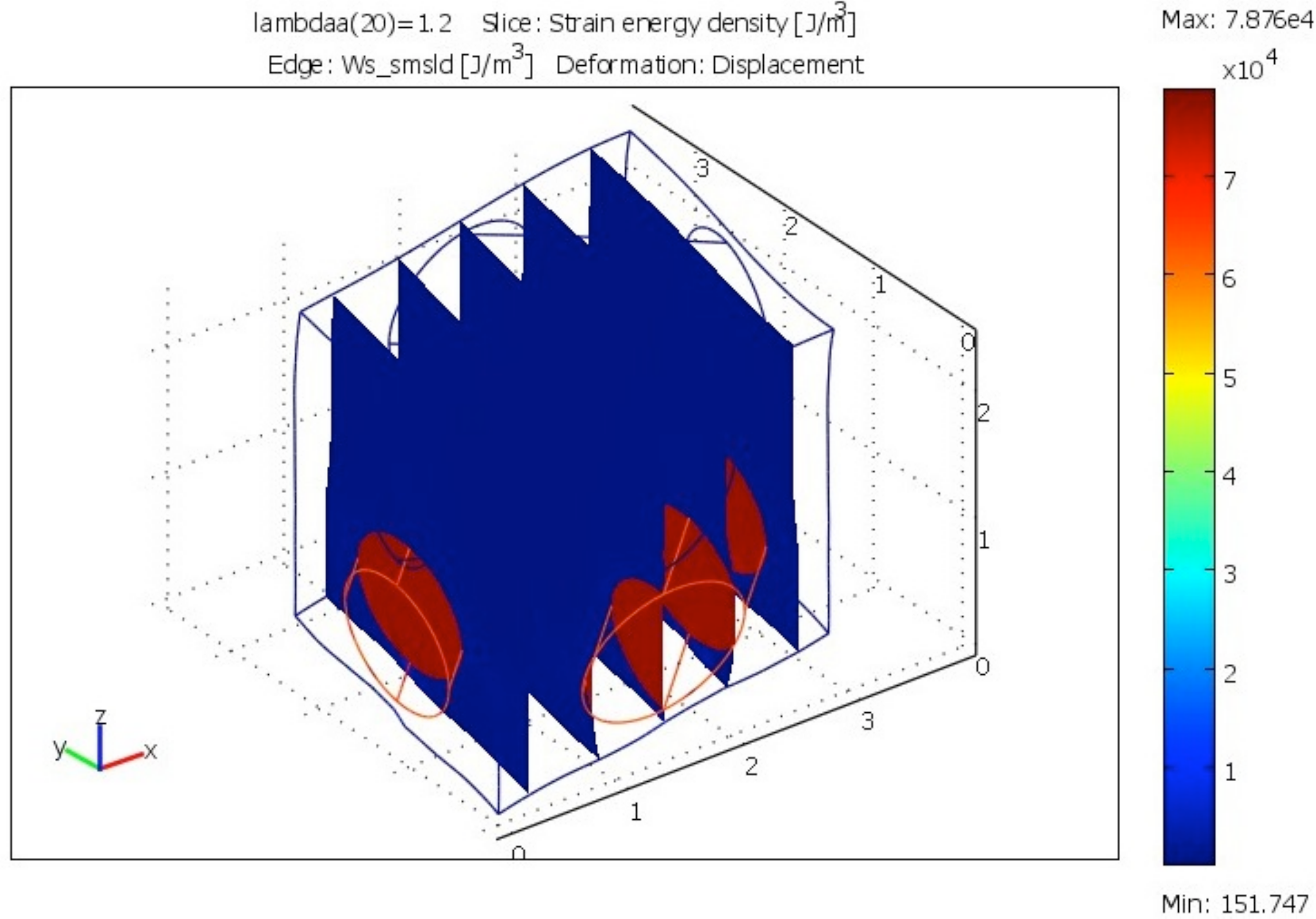}
  \quad
  (\emph{ii})\includegraphics[scale=0.28]{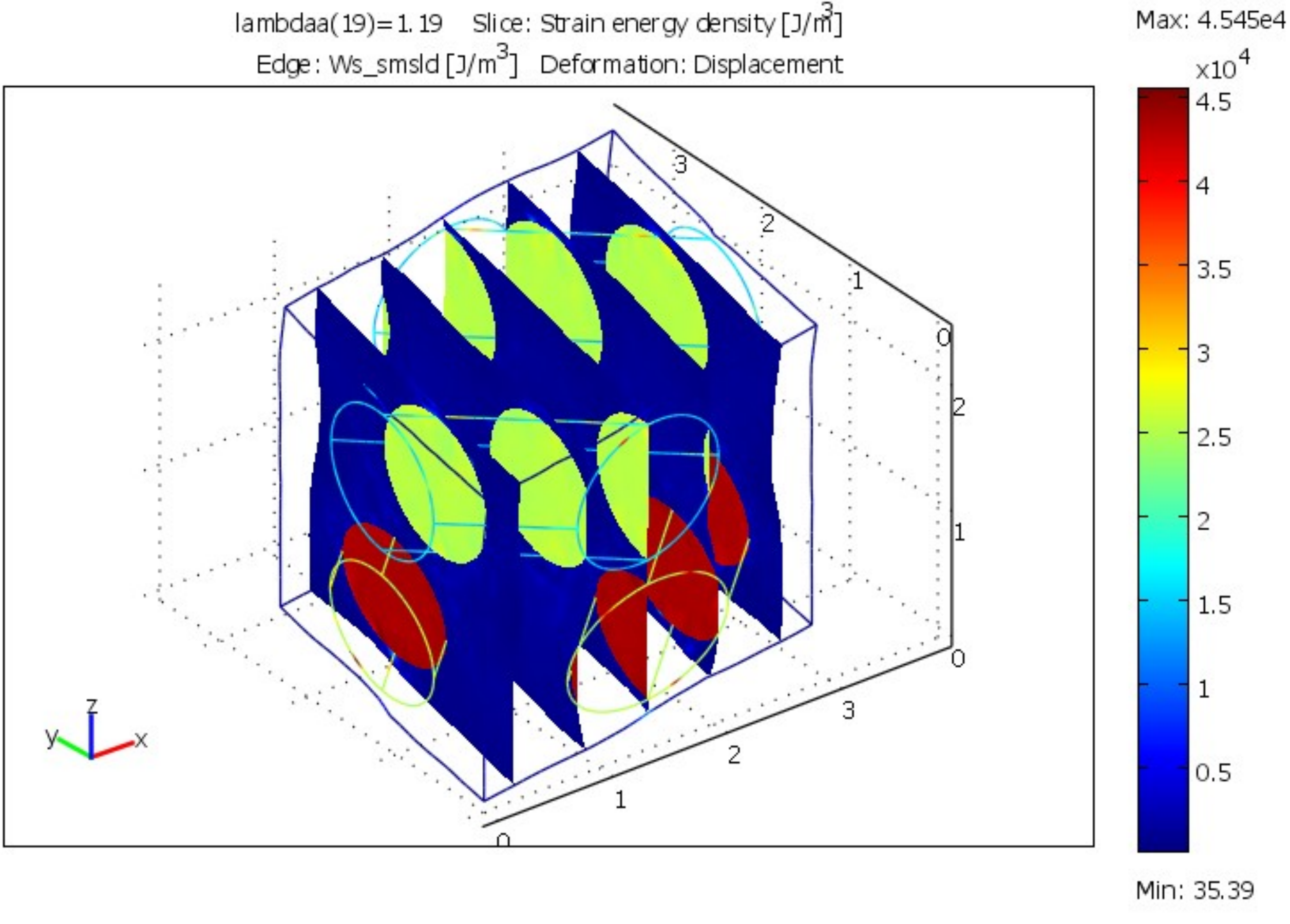}
  \caption{\footnotesize Deformed configuration of the unit cell during incline extension at $\rotangle=43^{\circ}$. Case-(\emph{i}) is shown in the left figure and case-(\emph{ii}) in the right one. The stored energy-density at several slices is depicted by the contour plots.}\label{drawing_incline_43}
\end{center}
\end{figure}
Deformed configurations of the periodic cell together with contour plots representing the stored energy are shown in Figs.~\ref{drawing_incline_21} and \ref{drawing_incline_43} for the inclined extensions at $\rotangle=21.5^{\circ}$ and $\rotangle=43^{\circ}$, respectively.
The severe macroscopic shear to which the cell is subjected can be appreciated.
In addition, the local shear resulting from the periodicity conditions is associated with the curvature of the cell faces.
In case-(\emph{ii}) the upper family of fibers which is subjected to compression can be easily distinguished due to the higher level of stored energy in comparison with the surrounding matrix. 
Contrarily, in case-(\emph{i}) the amount of the stored energy-density in the compressed fibers is identical to the one  in the surrounding EFM.

\begin{figure}[t]
\begin{center}
    (a)\includegraphics[scale=0.45]{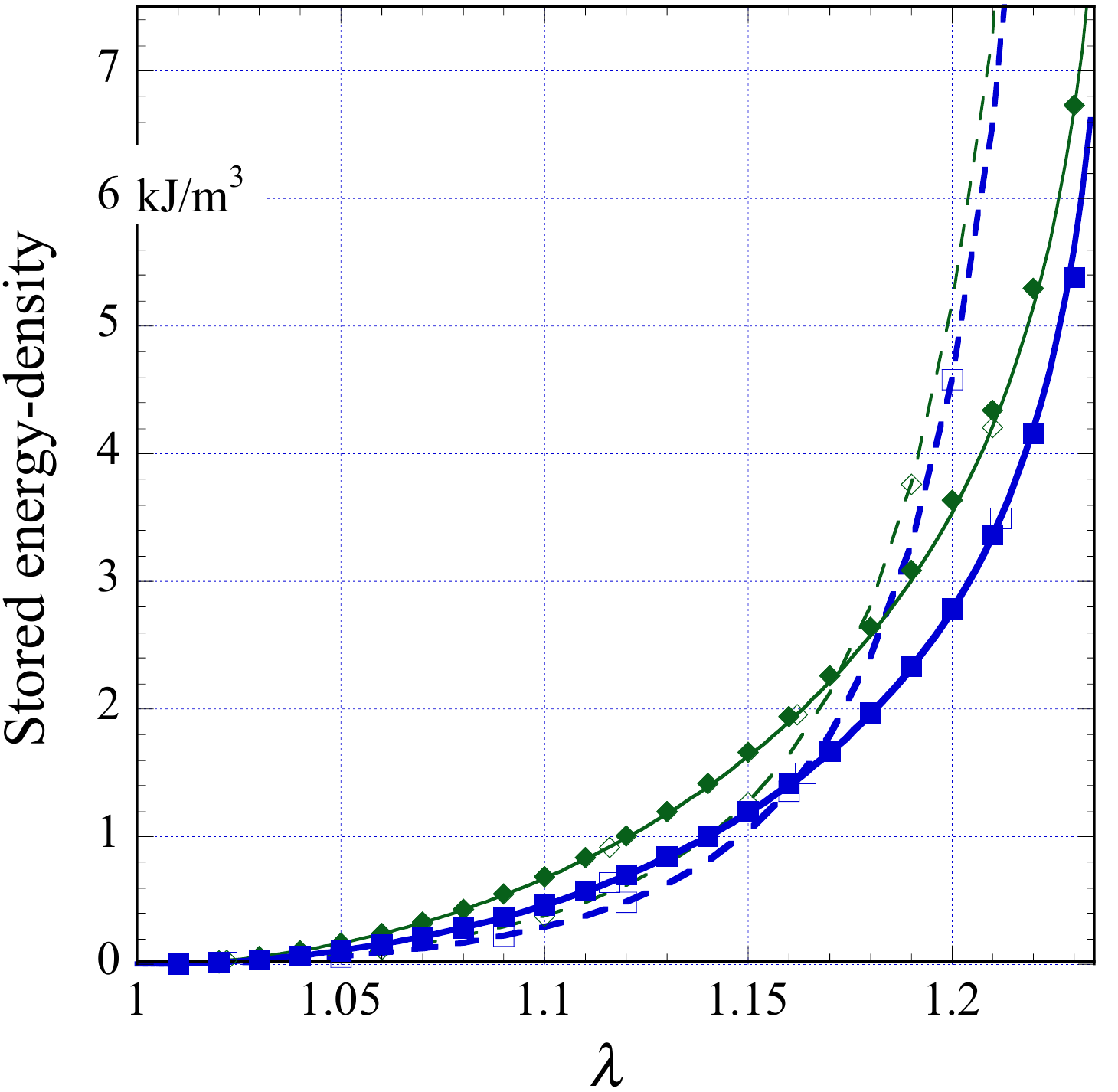}
  \quad\quad
  (b)\includegraphics[scale=0.45]{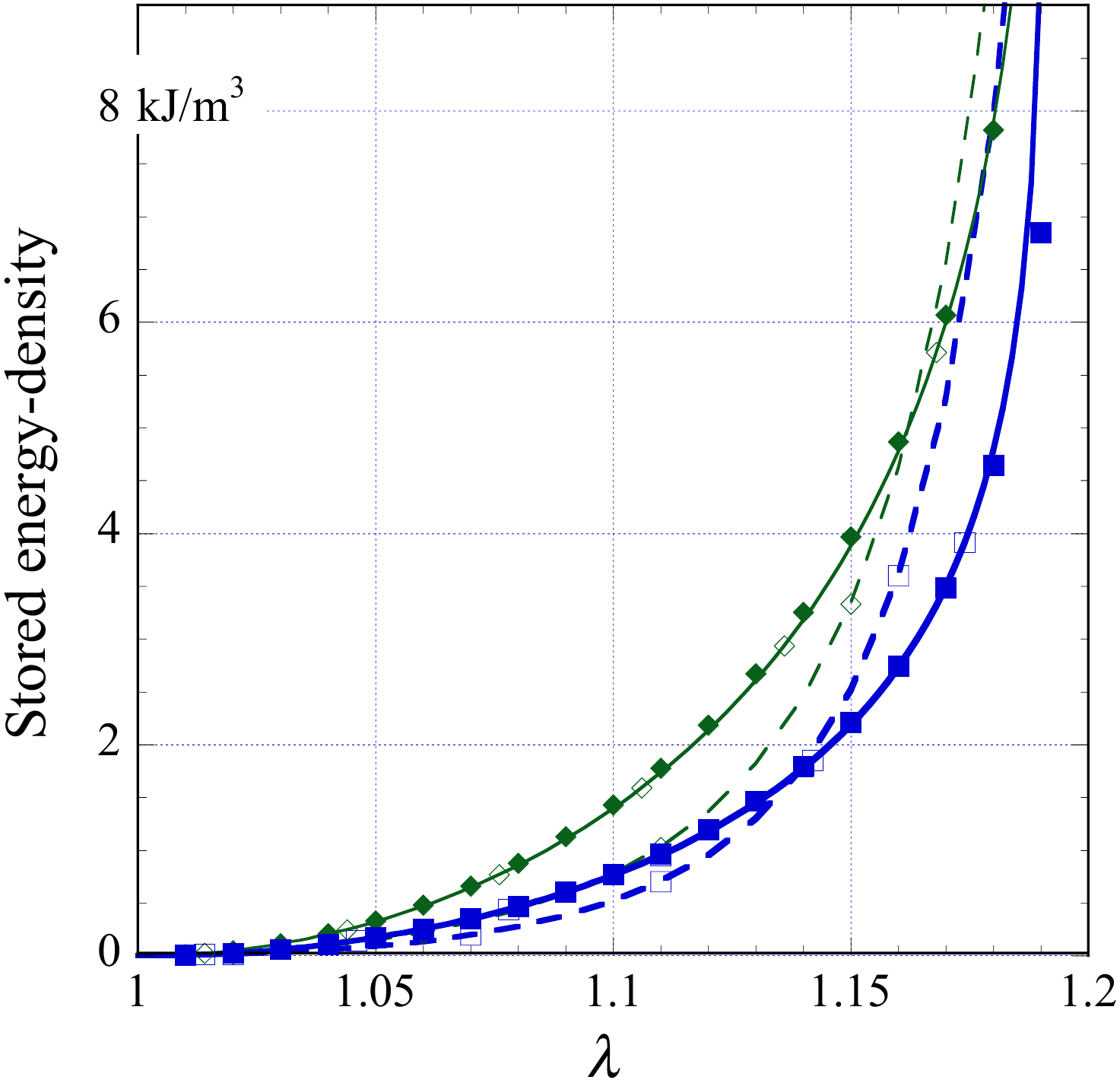}
  \caption{\footnotesize The stored elastic energy-density according to the NLC estimate (continuous curves), the periodic FE model (shaded marks), and the model of \citet{holz&etal05ajphcp} (dashed curves).
Figures (a) and (b) correspond to tensions in the $\rotangle=21.5^{\circ}$ and the $\rotangle=43^{\circ}$ directions, respectively.
Cases (\emph{i}) and (\emph{ii}) are signified by the blue square and the the green diamond marks, respectively.}
\label{stored_energy_incline}
\end{center}
\end{figure}
In Fig.~\ref{stored_energy_incline} the estimates for the stored energy-density according to the two micromechanics based models are shown.
Figs.~\ref{stored_energy_incline}(a) and (b) correspond to extensions at angles $\rotangle=21.5^{\circ}$ and $\rotangle=43^{\circ}$, respectively.
The NLC estimate is represented by the continuous curves, the periodic FE model by the dark marks, and we also depict the corresponding results according to the model of \citet{holz&etal05ajphcp} (dashed curves).
Case-(\emph{i}), with no contribution of the fibers in compression, is identified by blue square marks and case-(\emph{ii}), where their contribution is accounted for, by the green diamond marks.

Our first observation concerns the excellent agreement between the two micromechanics based models.
Thus, in all four cases the shaded marks that correspond to the FE simulations are lying on top of the continuous curves derived by the NLC estimate. 
This suggest that the close form NLC estimate accurately captures the response of the heterogeneous material.
Contrarily, while the general trend of the phenomenological model 
is reminiscent of the micromechanics models, the precise details are different.
This is because the different structure of the latter, and its inability to capture the local interactions between the collagen fibers and their surrounding. 
Particularly, the excessively stiff response predicted by the phenomenological model is due to the inherently embedded assumption that the fibers and the tissue are subjected to the same macroscopic deformation.

An interesting observation concerns the influence of the behavior of the fibers under compression.
The difference between cases (\emph{i}) and (\emph{ii}) is represented by the difference between the curves marked by blue squares and those marked by green diamonds, respectively.
We also computed the stored energy-density according to the phenomenological model for these two cases. 
Ones according to case-(\emph{i}) without the contribution of the compressed family (blue dashed curve with hollow square marks), and once according to case-(\emph{ii}) with the contribution of this family (green dashed curve with hollow diamond marks).

As anticipated, the curves corresponding to case-(\emph{i}) are laying beneath the ones for case-(\emph{ii}).
However, while initially the difference between the two pairs of curves increases, eventually the two are getting closer.
The reason is the rotation of the fibers towards the tension direction during the deformation.
Because of this, at some value of $\lambda$ the compressive strains in the compressed family start to decrease till at large enough stretch ratio they turn into tension \citep{gdb&shmu09jmps}.
The fact that the level of the stored energy in the compressed family is severely lower than the one in the tensed family can be seen in Figs.~\ref{drawing_incline_21}(\emph{ii}) and \ref{drawing_incline_43}(\emph{ii}).
It is important to note that according to the micromechanics based models the difference between the two cases is larger than the difference predicted by the phenomenological model.

\begin{figure}[t]
\begin{center}
    \includegraphics[scale=0.5]{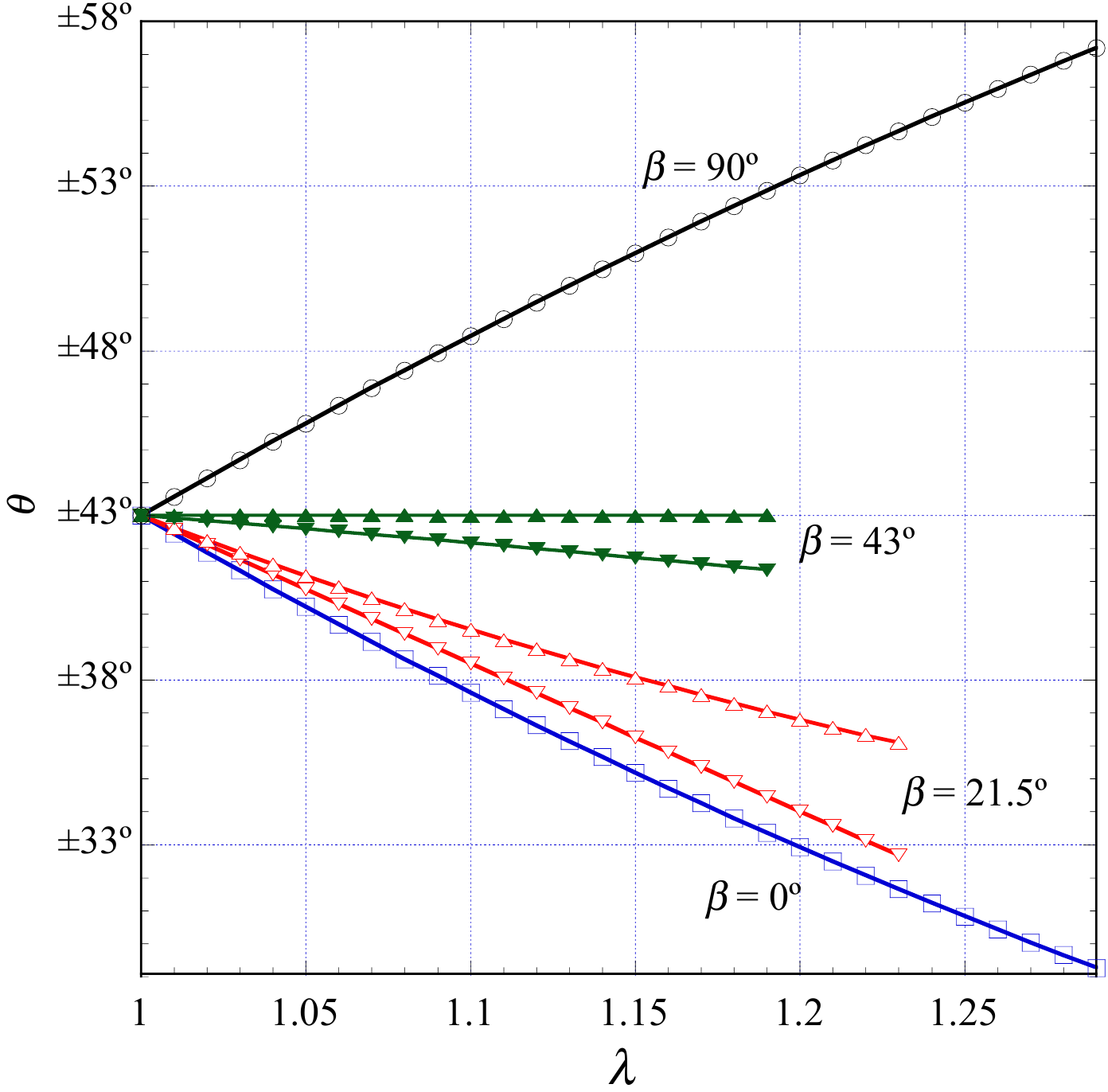}
  \caption{\footnotesize Evolution of the angle between the fiber directions and the principal orthotropic axis versus the tensile stretch ratio. 
The curves correspond to the analytical NLC estimate and the marks to the periodic FE model.}
\label{evolution}
\end{center}
\end{figure}
The rotation of the fibers, in terms of the evolution of the angle between the fiber and the principal orthotropic direction $X_{1}$ versus the principal stretch ratio is shown in Fig.~\ref{evolution} for the four different loadings shown in Figs.~\ref{drawing_principal_extensions}, \ref{drawing_incline_21} and \ref{drawing_incline_43}.
The continuous curves correspond to the results of the NLC estimate and the marks to the FE simulations.
Following the results of Appendix~\ref{laminate}, the analytical expression for the angle is simply
\begineq{fiber angle}
\theta=\arctan\left(\frac{\tan(\Theta-\rotangle)}{\lambda^{2}}\right)+\rotangle.
\end{equation}
There is an excellent agreement between the two micromechanics based models.

\begin{figure}[t]
\begin{center}
    (a)\includegraphics[scale=0.5]{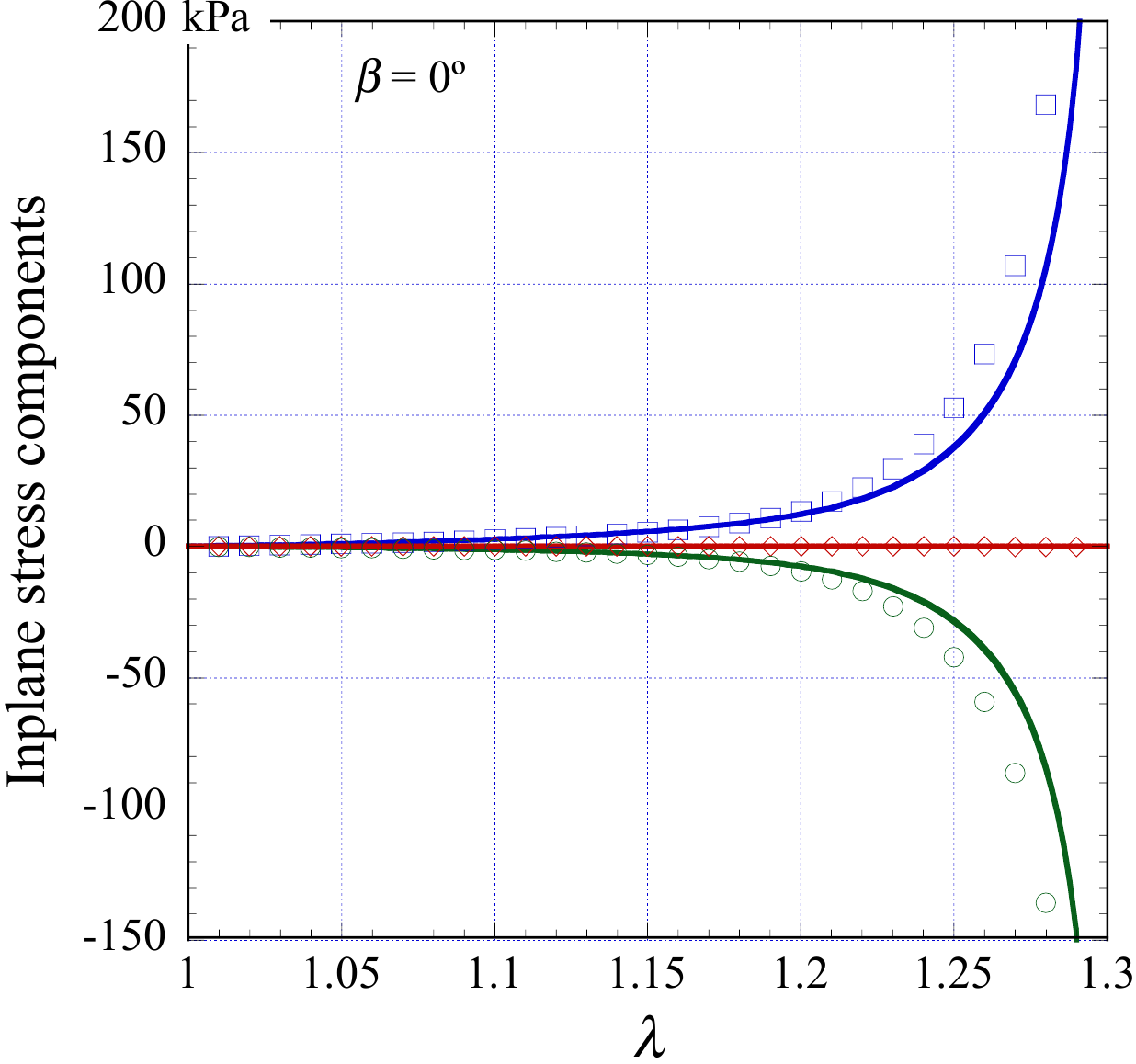}
  \quad\quad
  (b)\includegraphics[scale=0.50]{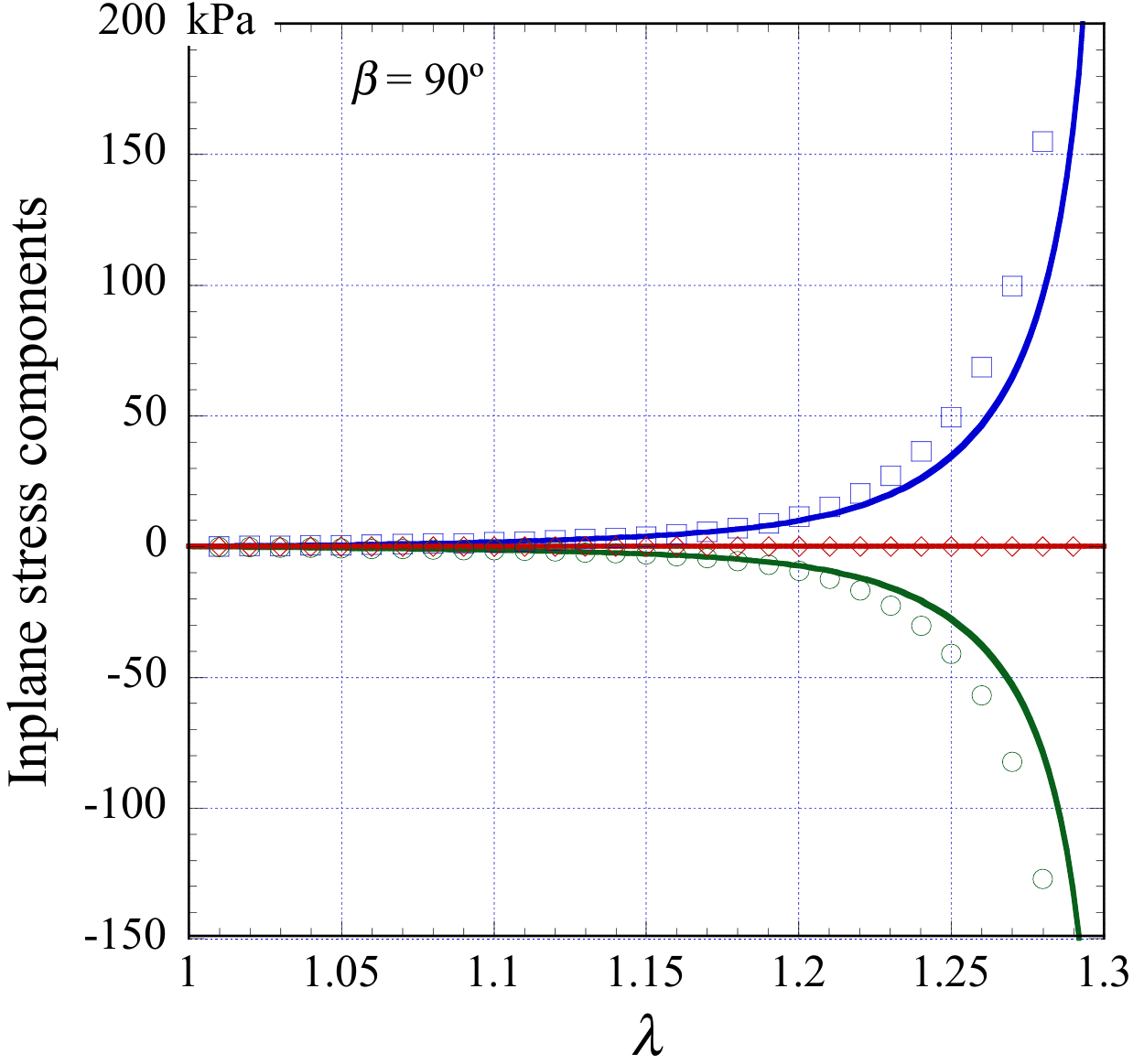}
  \caption{\footnotesize Deviatoric stresses in the tension direction (blue squares), in the transverse direction (green circles), and in-plane shear (red diamonds) versus the tensile stretch ratio.
The continuous curves correspond to the NLC estimate and the marks to the periodic FE model.
Figures (a) and (b) depict tensions along the $X_{1}$ and $X_{2}$ axes, respectively, respectively.}
\label{stress_0-90}
\end{center}
\end{figure}

To assess the ability of the NLC estimate to capture the behavior of the heterogeneous tissue we also compare the overall stresses determined via \Eq{stressTFF} with the corresponding stresses predicted by the finite element simulation according to \refeq{avgPf}.
In Fig.~\ref{stress_0-90} the predicted components of the deviatoric stress according to the different models are shown during tensions along the principal orthotropic axes of the tissue.
Figs.~\ref{stress_0-90}a and \ref{stress_0-90}b correspond to tensions along the in-plane orthotropic axes $X_{1}$ and $X_{2}$, respectively.
The corresponding deformed configurations are shown in Figs.~\ref{drawing_principal_extensions}a and \ref{drawing_principal_extensions}b, respectively.
Throughout, the curves correspond to the predictions of the NLC estimate, and the marks to the FE simulations.
The blue curves and square marks correspond to the stress in the tensile direction
(\eg~in the coordinate system of Fig.~\ref{UC}, $\cauchy_{11}^{D}$ for Fig.~\ref{stress_0-90}a and $\cauchy_{22}^{D}$ for Fig.~\ref{stress_0-90}b).
The compressive stresses in the perpendicular direction are represented by the green curves and the circular marks.
The in-plane shear stresses are represented by the red curves and the diamond marks.

The overall response of the tissue, which is dictated by the behavior of the collagen fiber, can be traced in these figures.
Initially, the contribution of the fiber is negligible and the slope of the stress-strain curve is very low.
At a stretch ratio of about 1.2 the contribution of the fibers becomes evident as the intensity of the stresses become larger than the shear moduli of the EFM. 
Subsequently, as the stretch ratio approaches 1.3, the stiff fibers dominate the response of the tissue and the slopes of the stress-strain curves increase dramatically.
In both figures, since the stretches are along the principal orthotropic axes, the macroscopic shear stresses vanish.
In the context of the present work, we emphasize the fine agreement between the analytical prediction according to the NLC estimate and the predictions according to the FE simulations.

\begin{figure}[t]
\begin{center}
    (a)\includegraphics[scale=0.5]{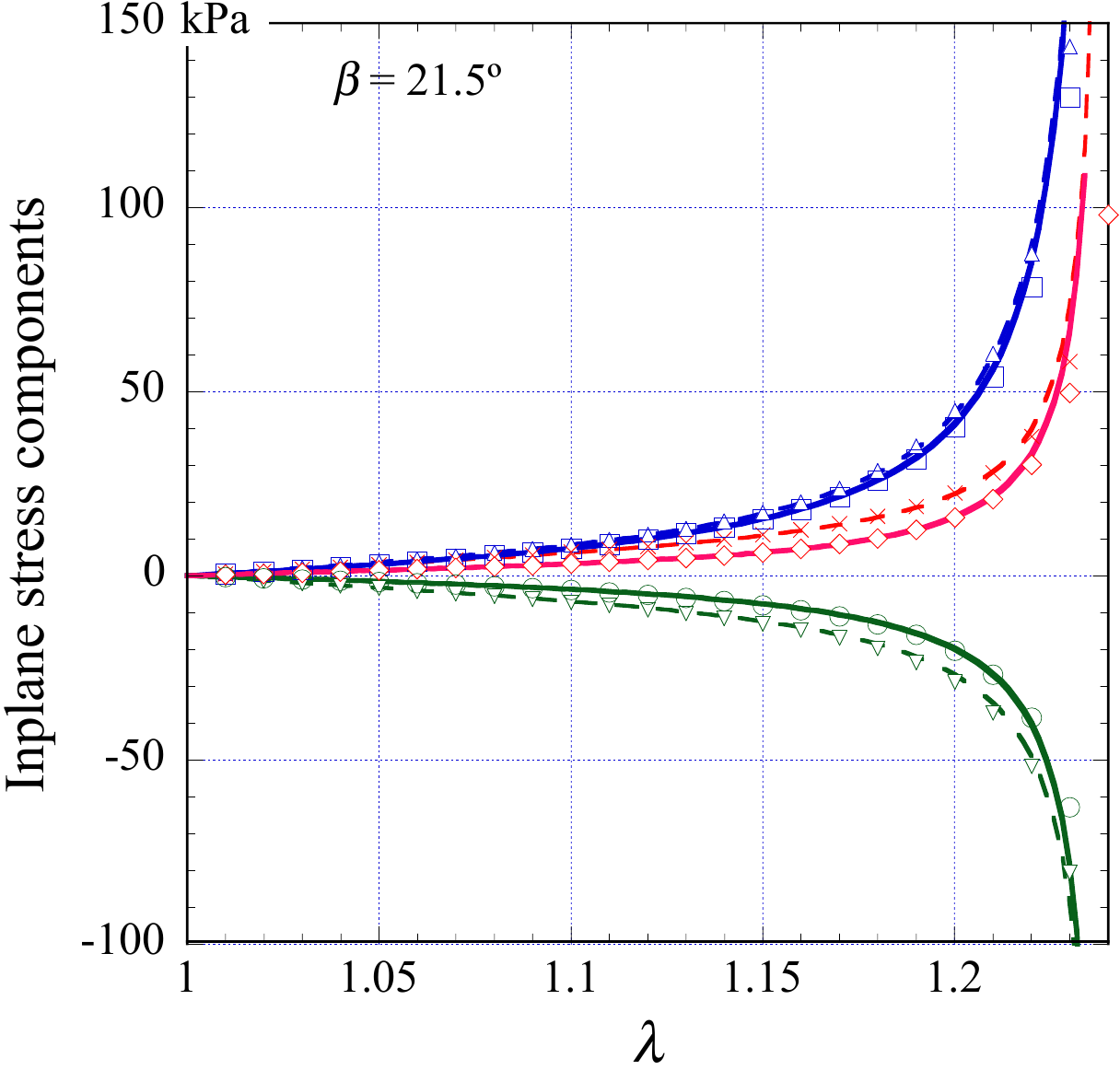}
  \quad\quad
  (b)\includegraphics[scale=0.50]{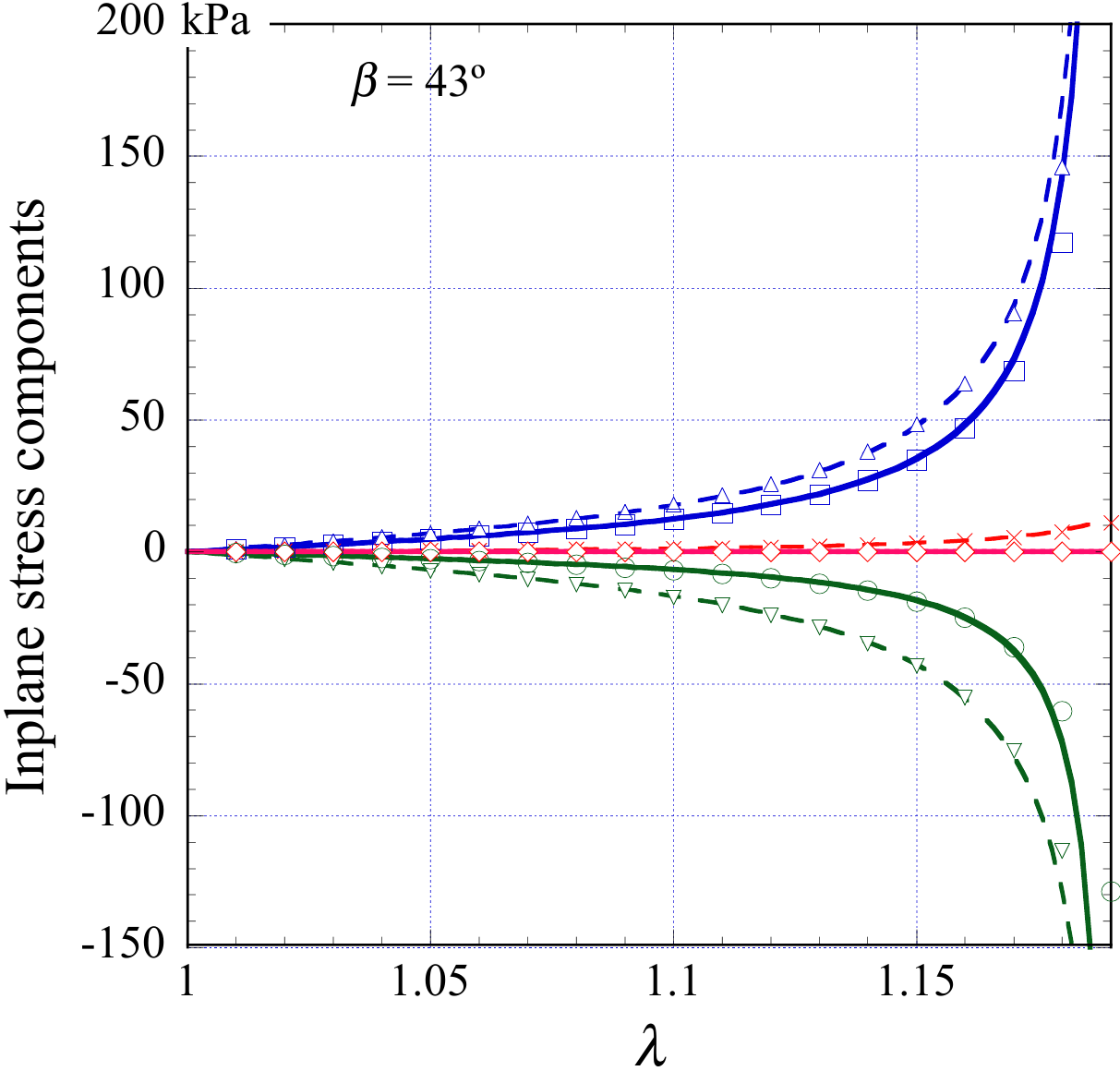}
  \caption{\footnotesize Deviatoric stresses in the tension direction (blue squares and upright triangles), in the transverse direction (green circles and inverted triangles), and in-plane shear (red diamonds and crosses) versus the tensile stretch ratio.
The curves correspond to the NLC estimate and the marks to the periodic FE model.
Case-(\emph{i}) corresponds to the continuous curves and the square, circle and diamond marks.
Case-(\emph{ii}) corresponds to the dashed curves and the upright triangle, inverted triangle and cross marks.}
\label{stress_21-43}
\end{center}
\end{figure}
The stresses developing due to incline extensions are shown in Fig.~\ref{stress_21-43}.
Figs.~\ref{stress_21-43}a and \ref{stress_21-43}b correspond to tensions at $\rotangle=21.5^{\circ}$ and $\rotangle=43^{\circ}$, respectively.
The corresponding deformed configurations are shown in Figs.~\ref{drawing_incline_21} and \ref{drawing_incline_43}, respectively.
Throughout, the curves correspond to the predictions of the NLC estimate, and the marks to the FE simulations.
Case-(\emph{i}) is represented by the continuous curves and the square, circle and diamond marks, and
Case-(\emph{ii}) by the dashed curves and the upright triangle, inverted triangle and cross marks.
The blue curves, square and upright triangle marks correspond to the stress in the tensile direction.
The perpendicular compressive stresses are represented by the green curves, circle and the inverted triangle marks.
The in-plane shear stresses are represented by the red curves, the diamond and the cross marks.

Once again, we observe that throughout the marks are laying on top of the curves. This indicates that there is an excellent agreement between the analytical NLC estimate and the corresponding FE simulations. 
The role of the fibers can be appreciated with regard to the overall response of the tissue.
In the case of tension at $\rotangle=21.5^{\circ}$, since the direction of the tension is closer to the direction of one of the families (the lower family in Fig.~\ref{UC}), the contribution of the fiber becomes evident at a faster rate and already at a stretch ratio of $\lambda=1.15$ the tissue stiffens.
Naturally, during a tension along one of the families ($\rotangle=43^{\circ}$) this phenomenon is further enhanced and the tissue stiffens already at $\lambda=1.1$.
An interesting observation concerns the shear stresses. 
When the extension is at $\rotangle=21.5^{\circ}$ large shear stresses develop in the tissue since the principal stretch is shifted from the principal orthotropic axes.
At $\rotangle=43^{\circ}$ the shear stress are much smaller since the load is carried by a single family and the rotation of the second family is negligible (see also Fig.~\ref{evolution}).
In particular, since in case-(\emph{i}) the extension is essentially along the preferred direction of the tissue, no shear stresses develop. 

\section{Concluding remarks}\label{conclusion}

Highly accurate and efficient methods for characterizing the mechanical response of biological tissues are required in modern analyses and modeling.
Micromechanics based approaches that enable to account for the behaviors of the individual constituents together with the histology of the tissue can provide realistic and versatile methods for characterizing the behaviors of broad classes of tissues.
For instance, if estimates for the behaviors of the collagen fibers and the EFM are available, by following this approach different parts of the arterial wall, different sections of the artery, and even different arteries can be modeled by modifying appropriate morphological parameters such as the volumetric content and the spatial arrangement of the fibers. 
Moreover, it is possible that many of the histological parameters will be measured by non-invasive techniques, and hence the need for large sets of experiments required for tuning up phenomenological models can be avoided. 

In this work we made use of micromechanics based estimates for soft connective tissues with one and two families of fibers.
Both analytical and numerical models were examined and their predictions for the overall response of a healthy human coronary artery were compared.
The analytical model, which is based on the variational estimate of \cite{gdb&shmu10jmps}, resulted in a close form  expression which is no more complicated than corresponding ad-hoc phenomenological models.
The numerical model is based on the homogenization of a periodic unit cell subjected to periodic boundary conditions.
The constitutive behaviors for the collagen fibers and the EFM were calibrated by comparison with a phenomenological model whose parameters were determined for the modeled tissue during circumferential and axial tensile tests.
Under inclined loadings, when one of the families is tensed and the other is compressed, we examined the behavior of the tissue when the compressed family either contributes to the overall response or not.
We find that under inclined loadings there is marked difference between the predictions of the phenomenological and the micromechanical estimates, where the former predicts a substantially stiffer response of the arterial wall.
Contrarily, for all loading conditions there is a fine agreement between the predictions for the overall stresses and evolution according to the two micromechanics based models.
This implies that the proposed close form analytical estimate adequately captures the response of connective tissues and the  interactions between the EFM and the two families of collagen fibers.


\appendix

\section{The effective behavior of an incompressible laminated medium}\label{laminate}
We follow the steps followed by \citet{gdb05jmps} and determine the effective strain energy-density function $\tilde\sedf^{(L)}(\bar\defgT)$ of a layered medium made out of alternating incompressible layers whose behaviors are characterized by the strain energy-density functions $\sedf^{(l)}(\defgT),~l=1,2$.
We assume that the thickness of the individual layers is at least an order of magnitude smaller than the thickness of the medium under consideration which, in turn, is smaller than its other dimensions.
This scale separation assumption implies that under homogeneous boundary condition of \Eq{homogeneous bc}
the local deformation gradients in the layers $\defgT^{(1)}$ and $\defgT^{(2)}$ are constants.
Accordingly, from \Eq{avgF}
\begineq{homogenized layer}
\bar\defgT=\sum_{l=1,2}\volfrac^{(l)}\defgT^{({l})}=\defgT_{0},
\end{equation}
where $\volfrac^{(l)}$ are the volume fractions of the layers composing the medium such that $\sum_{l=1,2}\volfrac^{(l)}=1$.
The continuity of the displacements across the interfaces between two adjacent layers are
\begineq{displacement continuity}
(\defgT^{(1)}-\defgT^{(2)})\hat{\mathbf{M}}_{L}=\mathbf{0},
\end{equation}
where $\hat{\mathbf{M}}_{L}$ is any unit vector in the layers plane such that $\hat{\mathbf{M}}_{L}\cdot\refnormalT_{L}=0$ with $\refnormalT_{L}$ being the unit vector normal to the layers plane.
Thus, only the components of the normal projections of the local deformation gradients $\defgT^{(l)}\refnormalT_{L}$ may differ.
Specifically, in the coordinate system of Fig.~\ref{UC}a, only the components $\defg_{13}$, $\defg_{23}$ and $\defg_{33}$ may differ.
If we assume that the anti-plane shear components are vanishingly small (\ie~$(\defgT\refnormalT_{L})\cdot\hat{\mathbf{M}}_{L}=0$ for any vector $\hat{\mathbf{M}}_{L}$, or in the chosen coordinate system $\defg_{13}=\defg_{23}=0$), then in each layer $(\defgT^{(l)}\refnormalT_{L})\cdot\refnormalT_{L}=(\bar\defgT\refnormalT_{L})\cdot\refnormalT_{L}$
since due to the incompressibility constraint these depend on the continuous in-plane components (in the chosen coordinate system $\defg_{33}^{(l)}=\bar\defg_{33}$).
In summary, Eqs.~\eqref{homogenized layer}, \eqref{displacement continuity} and the incompressibility assumption lead to the conclusion that under vanishingly small anti-plane shear loads
\begineq{identical defg}
\defgT^{({l})}=\bar\defgT,~l=1,2.
\end{equation}
Consequently, the solution of the optimization problem \Eq{Wmin} is trivial, and the effective energy-density function of the layered medium is nothing but the volume average of the constituents energy-density functions, namely
\begineq{layered sedf}
\tilde\sedf^{(L)}(\bar\defgT)=\sum_{l=1,2}\volfrac^{(l)}\sedf^{({l})}(\bar\defgT).
\end{equation}
We finally note that the above developments are independent of the precise constitutive behaviors of the phases composing the laminate. 
Moreover, the above arguments can be trivially extended to laminated media with more than two distinct phases (\ie~$l>2$).
This last observation may become useful if a distribution of the fibers is accounted for.

\renewcommand{\baselinestretch}{1.2}
\small 

\end{document}